\documentclass[prd,%
preprint,%
tightenlines,showpacs,%
nofootinbib,%
superscriptaddress,%
amsfonts,amsmath]{revtex4}

\begin{document}

\title{Counting the degrees of freedom of generalized Galileons}

\author{C\'edric~Deffayet} \email{deffayet@iap.fr}
\affiliation{${\mathcal{G}}{\mathbb{R}}\varepsilon{\mathbb{C}}{\mathcal{O}}$,
Institut d'Astrophysique de Paris,\\
CNRS, UMR 7095, et Sorbonne Universit\'es,
UPMC Univ Paris 6,\\
98bis boulevard Arago, F-75014 Paris, France}
\affiliation{IH\'ES, Le Bois-Marie, 35 route de Chartres,
F-91440 Bures-sur-Yvette, France}

\author{Gilles~\surname{Esposito-Far\`ese}} \email{gef@iap.fr}
\affiliation{${\mathcal{G}}{\mathbb{R}}\varepsilon{\mathbb{C}}{\mathcal{O}}$,
Institut d'Astrophysique de Paris,\\
CNRS, UMR 7095, et Sorbonne Universit\'es,
UPMC Univ Paris 6,\\
98bis boulevard Arago, F-75014 Paris, France}

\author{Dani\`ele A.~Steer} \email{steer@apc.univ-paris7.fr}
\affiliation{AstroParticule \& Cosmologie,
UMR 7164-CNRS, Universit\'e Denis Diderot-Paris 7,
CEA, Observatoire de Paris,
10 rue Alice Domon et L\'eonie
Duquet, F-75205 Paris Cedex 13, France}
\affiliation{${\mathcal{G}}{\mathbb{R}}\varepsilon{\mathbb{C}}{\mathcal{O}}$,
Institut d'Astrophysique de Paris,\\
CNRS, UMR 7095, et Sorbonne Universit\'es,
UPMC Univ Paris 6,\\
98bis boulevard Arago, F-75014 Paris, France}
\affiliation{LAPTH, Universit\'e Savoie Mont Blanc, CNRS,
B.P.110, F-74941 Annecy-le-Vieux Cedex, France}

\begin{abstract}
We consider Galileon models on curved spacetime, as well as the
counterterms introduced to maintain the second-order nature of
the field equations of these models when both the metric and the
scalar are made dynamical. Working in a gauge invariant
framework, we first show how all the third-order time derivatives
appearing in the field equations ---~both metric and scalar~---
of a Galileon model or one defined by a given counterterm can be
eliminated to leave field equations which contain at most
second-order time derivatives of the metric and of the scalar.
The same is shown to hold for arbitrary linear combinations of
such models, as well as their k-essence-like/Horndeski
generalizations. This supports the claim that the number of
degrees of freedom in these models is only 3, counting 2 for the
graviton and 1 for the scalar. We comment on the arguments given
previously in support of this claim. We then prove that this
number of degrees of freedom is strictly less that 4 in one
particular such model by carrying out a full-fledged Hamiltonian
analysis. In contrast to previous results, our analyses do not
assume any particular gauge choice of restricted applicability.

\end{abstract}

\date{July 16, 2015}


\maketitle

\section{Introduction}
\label{Sec1}

Scalar-tensor theories are widely used in cosmology and
extensions of general relativity, with applications ranging from
inflation to the late-time observed acceleration of the Universe,
and tests of gravitation. Motivated in part by the ability of
some of these theories to give an alternative to dark energy,
there has recently been renewed interest in the delineation of
theories in which ---~besides the two degrees of freedom of a
massless graviton~--- there is only one propagating degree of
freedom (d.o.f.) stemming from the scalar. Along this line, an
important result was achieved by Horndeski who classified all
scalar-tensor theories in 4 dimensions having field equations
(both for the metric and the scalar) with derivatives of order
less than or equal to two \cite{Horndeski}. Similarly,
Ref.~\cite{Nicolis:2008in} (see also \cite{Fairlie} for earlier
works) introduced, on flat spacetime and for an arbitrary number
of dimensions $D$, a set of scalar theories with field equations
exactly of second order: the Galileons. These theories were later
``covariantized'', i.e., put on arbitrary curved spacetime with a
dynamical metric, while maintaining the second-order nature of
the scalar field equation, as well as yielding metric field
equations of the same order
\cite{Deffayet:2009wt,Deffayet:2009mn}. This covariantization
procedure involves a non-minimal coupling between the curvature
and the scalar in the form of very specific counterterms able to
remove all higher derivatives from the field equations. Indeed a
minimal covariantization of the original Galileon of
Ref.~\cite{Nicolis:2008in} (i.e., the mere replacement of partial
derivatives by covariant derivatives in the action) was shown to
lead to third-order derivatives in the field
equations.\footnote{We refer to such models as ``minimally
covariantized'' Galileons, to stress the difference with the
``covariantized'' Galileons of
ref.~\cite{Horndeski,Deffayet:2009wt,Deffayet:2009mn} which
contain non-minimal coupling to curvature in the form of the
counterterms mentioned above.} Another relevant work is that of
\cite{Deffayet:2011gz}, which classified all scalar theories
having equations of motion of order less than or equal to two on
a flat spacetime of arbitrary dimension, and then covariantized
these theories. It was shown that the original flat-space time
Galileons~\cite{Nicolis:2008in}, their flat spacetime
generalizations \cite{Deffayet:2011gz} as well as their
covariantization
\cite{Deffayet:2009wt,Deffayet:2009mn,Deffayet:2011gz} (with the
meaning above) belong, for a spacetime with 4 dimensions, to the
set of Horndeski (as they should according to Horndeski theorem)
\cite{Deffayet:2011gz,Kobayashi:2011nu}. These theories were also
generalized to the case of multiscalars and $p$-forms
\cite{Deffayet:2010zh,Padilla:2010de,Padilla:2010tj,Hinterbichler:2010xn,Trodden:2011xh,Sivanesan:2013tba,Padilla:2012dx,Deffayet:2013tca,Ohashi:2015fma}.

Having covariant second-order field equations is \textit{a
priori} enough, once diffeomorphism invariance is taken into
account, to have just 3 propagating degrees of freedom in vacuum
(counting 2 for the metric and 1 for the scalar), and to put the
theory on the safe side as far as Ostrogradski's type of
instability is concerned \cite{Ostrogradski,Woodard:2006nt}.
However, to the best of our knowledge, a proper Hamiltonian
counting of degrees of freedom in these theories, including the
ones contained in the metric, has so far not been carried out
(the flat spacetime limit has been analyzed in
Refs.~\cite{Zhou:2010di,Sivanesan:2011kw}, while some other
references start from a gauge-fixed action in which the gauge
invariance has not been properly fixed or is explicitly broken
\cite{Gleyzes:2014dya,Gleyzes:2014qga,Lin:2014jga,Gao:2014fra}).
In fact, the Hamiltonian analysis is complicated by the kinetic
mixing (or braiding to use the wording of \cite{Deffayet:2010qz})
between the scalar and the metric, and one aim of the present
work is to provide a first step towards a proper Hamiltonian
treatment of Galileon-like theories.

A second motivation stems from the work of
\cite{Gleyzes:2014dya,Gleyzes:2014qga}, building on the earlier
works of \cite{Chen:2012au,Zumalacarregui:2013pma}, suggesting
that despite the presence of higher derivatives in the field
equations of minimally covariantized Galileons, the number of
propagating degrees of freedom can still be only three due to the
presence of some hidden constraint in the theory. (In minimally
covariantized Galileons and related models, as stressed in
\cite{Deffayet:2009wt,Deffayet:2009mn}, the field equations for
the metric are second order for the metric but contain
third-order derivatives of the scalar, and conversely the scalar
field equation is second order for the scalar but third-order for
the metric.) This appears, of course, to be a perfectly
legitimate possibility and it is not hard to build some simple
examples with such a property (see e.g.~\cite{Gabadadze:2012tr}).
However, so far, the arguments given in favor of this happening,
as proposed in
Refs.~\cite{Gleyzes:2014dya,Gleyzes:2014qga,Zumalacarregui:2013pma},
do not appear to us to be entirely conclusive.

The reasons are the following. First, the Hamiltonian analysis of
the corresponding theory has only been carried starting from a
gauge fixed action
\cite{Gleyzes:2014dya,Gleyzes:2014qga,Lin:2014jga} where (i)~the
scalar gradient $\nabla_\mu \varphi$ is assumed to give the time
flow direction, and (ii)~this gauge fixing is not included in the
Hamiltonian (i.e., the gauge is explicitly broken to start with).
This gauge, usually referred to as the ``unitary gauge'', hides
all the dynamics of the scalar and it is easy to see that it
eliminates all third time derivatives in the field equations (see
Refs.~\cite{Deffayet:2009wt,Deffayet:2009mn} and also
Sec.~\ref{Sec3} below). In this sense, it is perhaps not
surprising that working in this gauge, one finds less degrees of
freedom than those expected from an Ostrogradski-type of
reasoning. Furthermore, this unitary gauge choice is obviously
only possible if the scalar gradient is everywhere time-like (or
at least time-like in the vicinity of some would-be Cauchy
surface), a situation which only covers rather limited subset of
all possibilities. Indeed, it does not allow one to say anything
about the Cauchy problem when, on the Cauchy surface, the scalar
has a gradient which is not always time-like ---~a perfectly
legitimate choice of initial condition. For instance, a physical
situation of major importance of this kind is that of a static
and spherically symmetric background, since the hypersurfaces
$\varphi = \text{const.}$ are not spacelike and cannot be chosen
as initial value surfaces. Second, it is well known, e.g.~when
considering Maxwell theory, that analyzing the d.o.f.~content of
the theory in a given gauge can be very misleading, in particular
when the gauge is explicitly broken to start with. Finally, the
covariant reasoning given in \cite{Gleyzes:2014qga} (analogous to
the one of \cite{Zumalacarregui:2013pma}) appears to us to be
incomplete if not incorrect. Indeed, there it is stated that
taking an appropriate trace of the metric field equations (which
are known to contain third-order time derivatives of the scalar)
enables one to extract the third time derivatives of the scalar
in terms of second derivatives ---~and then, inserting this back
in the metric field equations, gives a second-order system (and
similarly with the scalar field equation). This seems to omit the
fact that in this way one can at best eliminate from the field
equations {\it all but one} third-order time derivatives of the
scalar (and similarly for the metric): the reason is that the
trace of the metric field equations is itself a field equation
which must still be solved, and which still contains a
third-order time derivative. Hence, in contrast to the claims in
\cite{Gleyzes:2014qga,Zumalacarregui:2013pma}, the covariant
procedure outlined in those papers appears not to lead to a
complete set of equations in which all third-order time
derivatives have been eliminated.

Here we will reexamine these issues and argue, in two different
ways, that minimally covariantized Galileons indeed propagate
less degrees of freedom than expected from the third-order nature
of the field equations. Throughout we work in a totally
gauge-invariant framework. This paper is organized as follows. In
Sec.~\ref{Sec2} we show how the system of $\frac{D(D+1)}{2}+1$
field equations\footnote{This number $\frac{D(D+1)}{2}+1$ will be
quoted several times in the present paper. In $D=4$ dimensions,
it simply reduces to the usual $10+1 = 11$ field equations of the
metric and the scalar field.} of the theory considered ---~namely
all minimally covariantized Galileons and independently all the
counterterms, as well as any linear combination of them and their
Horndeski-like generalizations~--- can be reduced to an
equivalent system with only second-order time derivatives (using,
however, a very different procedure from the one given in
Refs.~\cite{Gleyzes:2014qga,Zumalacarregui:2013pma}). In
Sec.~\ref{Sec3}, we present a Hamiltonian analysis of one
particular theory in the minimally covariantized Galileon family,
namely the simplest non trivial one (in the sense that its field
equations do contain third-order derivatives), to show that the
number of constraints is sufficient to yield less than 4
propagating degrees of freedom. A last section gives our
conclusions.

\section{Removal of third time derivatives in the general case}
\label{Sec2}

In a spacetime with metric $g_{\mu \nu}$ in any dimension $D$,
Ref.~\cite{Deffayet:2009mn} introduced the generalized Galileon
Lagrangians ${\cal L}_{(n+1,p)}$. These involve a total of $2n$
derivatives acting on a product of $n+1$ scalar fields $\varphi$,
and $p$ Riemann tensors $R_{\lambda\mu\nu\rho}$. The Lagrangians
with $p=0$, the ${\cal L}_{(n+1,0)}$, are the ``minimally
covariantized'' Galileons, whereas the Lagrangians with $p\neq 0$
are called the ``counterterms''.

For a given $n$, it was shown that ---~up to an irrelevant global
factor~--- there exists a {\it unique} linear combination
\begin{equation}
\sum_{p=0}^{p_{max}} \mathcal{C}_{(n+1,p)} {\cal L}_{(n+1,p)}
\label{CovG}
\end{equation}
such that all field equations are of second order. These are the
``covariantized'' Galileons. Here
\begin{equation}
p_{max} = \left\lfloor\frac{n-1}{2}\right\rfloor
\label{pmax}
\end{equation}
is the integer part of $\frac{n-1}{2}$, and the constant
coefficients $\mathcal{C}_{(n+1,p)}$ take a very specific form
which may be found in Eq.~(37) of \cite{Deffayet:2009mn}.

Conversely, any {\it other} linear combination (for instance,
each of these Lagrangians ${\cal L}_{(n+1,p)}$ individually when
$p_{max}>0$) {\textit{does}} yield third derivatives in its field
equations. More specifically, the scalar field equation contains
third time derivatives of the metric tensor, and the Einstein
equations contain the third time derivative of the scalar field.
To be able to compute the time evolution, it seems thus necessary
to specify more initial data on a Cauchy surface, and one expects
the existence of more degrees of freedom than just a single
scalar field and the two helicities of the graviton.

Although higher-order field equations indeed generically lead to
extra degrees of freedom (which are even generically ghost modes
\cite{Ostrogradski,Woodard:2006nt}, implying the instability of
the theory), specific examples show that this is not always the
case. There may, for instance, exist extra constraints (related
or not to some hidden gauge symmetry) which kill some of the
modes. Or the few equations involving third (or higher) time
derivatives may actually be obtained by differentiating, with
respect to time, some independently known second-order field
equations. In this case, extra initial data are not necessarily
needed on the Cauchy surface. An elegant toy-model of this kind
was presented in Sec.~7.1 of \cite{Gabadadze:2012tr}, and a
similar result applies to ``mimetic dark matter''
\cite{Chamseddine:2013kea}.

The aim of this section is to show that the field equations of
the generalized Galileon Lagrangians ${\cal L}_{(n+1,p)}$ are of
this second kind: all the third time derivatives can be obtained
by deriving independently known second-order field equations, and
can therefore be removed from all field equations. This supports
the main claims of
Refs.~\cite{Gleyzes:2014dya,Gleyzes:2014qga,Lin:2014jga},
although our procedure differs from theirs, and does not suffer
{}from the problems mentioned in the Introduction. In particular,
we will not fix any gauge in our derivation.

Thus we consider the $D$-dimensional theories defined by the
action
\begin{equation}
S = S_\text{EH} + S_\text{Gal}
\end{equation}
in curved spacetime, where $S_\text{EH} = \int d^Dx \sqrt{-g} R$
is the Einstein-Hilbert action\footnote{Throughout this paper, we
use the sign conventions of Ref.~\cite{MTW}, and notably the
mostly-plus signature.}, without any factor $c^3/16\pi G$ to
simplify our discussion, and where
\begin{equation}
S_\text{Gal} = \int d^Dx \sqrt{-g}
\left(\sum_{n,p} k_{(n,p)} {\cal L}_{(n,p)}\right),
\label{SGal}
\end{equation}
with {\it arbitrary} constant coefficients $k_{(n,p)}$ ---~and
hence not the specific $\mathcal{C}_{(n,p)}$ discussed above. (We
will furthermore consider the Horndeski-like generalization of
these theories at the end of the section.) We also define the
Galileon energy-momentum tensor as
\begin{equation}
T^{\mu \nu} \equiv
\frac{1}{\sqrt{-g}} \, \frac{\delta S_\text{Gal}}{\delta g_{\mu\nu}},
\label{defTmunu}
\end{equation}
without any factor 2, so that Einstein's equations (i.e., the
field equations for the metric) simply read $G^{\mu\nu} =
T^{\mu\nu}$, where $G^{\mu\nu}$ is the Einstein tensor. Finally,
we define ${\cal E} \equiv \delta S_\text{Gal}/\delta \varphi$,
so that the scalar field equation reads ${\cal E} = 0$.

Note that this scalar field equation is a consequence of
Einstein's equations since, because of the diffeomorphism
invariance of action $S$, it follows that
\begin{equation}
\varphi^{\nu}\,{\cal E} =
-2\nabla_\mu\left(G^{\mu\nu}-T^{\mu \nu}\right),
\label{conservTmunu}
\end{equation}
where $\varphi^{\nu} \equiv \nabla^\nu\varphi$ denotes the
covariant derivative of the scalar field (without writing any
semicolon, to simplify; we shall also write
$\varphi_{\mu\nu\dots} \equiv
\nabla\dots\nabla_\nu\nabla_\mu\varphi$ in the following).
Independently of the diffeomorphism-invariance argument, this can
also be checked explicitly for the general Lagrangians ${\cal
L}_{(n,p)}$ or for particular cases \cite{Deffayet:2009wt}.
Therefore, if we manage to prove that Einstein's equations can be
recast as a set of second-order differential equations (with
respect to time), Eq.~(\ref{conservTmunu}) shows that the third
time derivatives of the metric tensor entering the scalar field
equation ${\cal E} = 0$ should not pose more problems than in the
toy-model of Ref.~\cite{Gabadadze:2012tr}. It should be noted
that at the spacetime points where $\varphi^{\nu}$ happens to
vanish, $\cal E$ can no longer be extracted from
(\ref{conservTmunu}). However, it is easy to see that in all
field equations, third derivatives are always multiplied by a
gradient $\varphi^{\nu}$ (and even several of them). Therefore,
at the points where $\varphi^{\nu} = 0$, all field equations are
at most of second order, and do not pose any problem. In addition
to the above argument based on Eq.~(\ref{conservTmunu}), we will
actually prove an even stronger result below: It is also possible
to recast the scalar field equation itself as a second-order one
(with respect to time), by combining it with the time derivative
of another linear combination of Einstein's equations.

\subsection{Example: ${\cal{L}}_{(4,0)}$ in $D=4$ dimensions}
\label{SubSec2A}

Before attacking the general case of ${\cal L}_{(n,p)}$ which is
rather technically involved, we begin in this subsection by
illustrating how the steps work in $D=4$ dimensions, focusing on
the simplest non-trivial Galileon action, namely $S_{\rm{Gal}} =
\int d^4 x \sqrt{-g} {\cal{L}}_{(4,0)}$.

The Lagrangian ${\cal{L}}_{(4,0)}$, which is also the subject of
the Hamiltonian analysis of Sec.~\ref{Sec3}, is given by
\cite{Deffayet:2009mn}
\begin{eqnarray}
{\cal L}_{(4,0)} &=& \left(\Box \varphi\right)^2
\left(\varphi_{\mu}\varphi^{\mu}\right)
-2 \left(\Box \varphi\right)\left(\varphi_{\mu}
\varphi^{\mu\nu}\varphi_{\nu}\right)
- \left(\varphi_{\mu\nu}\varphi^{\mu\nu}\right)
\left(\varphi_{\rho}\varphi^{\rho}\right)
+2 \left(\varphi_{\mu}\varphi^{\mu\nu}
\varphi_{\nu\rho}\varphi^{\rho}\right)
\nonumber\\
&=& - \varepsilon^{\mu_{\vphantom{()}1}
\mu_{\vphantom{()}3} \mu_{\vphantom{()}5} \alpha}
\varepsilon^{\mu_{\vphantom{()}2}
\mu_{\vphantom{()}4} \mu_{\vphantom{()}6}}{}_{\alpha}
\varphi_{\mu_{\vphantom{()}1}} \varphi_{\mu_{\vphantom{()}2}}
\varphi_{\mu_{\vphantom{()}3} \mu_{\vphantom{()}4}}
\varphi_{\mu_{\vphantom{()}5} \mu_{\vphantom{()}6}},
\label{D:L4}
\end{eqnarray}
where $ \varepsilon^{\mu\nu\rho\sigma}$ is the Levi-Civita (fully
antisymmetric) tensor in 4 dimensions [see Eq.~(\ref{epsilon})
for its definition in $D$ dimensions]. Its stress-energy tensor
is given by
\begin{eqnarray}
T^{\mu \nu}_{(4,0)}
&=&
\left( \frac{1}{2} g^{\mu \nu} \varepsilon^{\mu_1 \mu_3 \mu_5 \alpha}
\varepsilon^{\mu_2 \mu_4 \mu_6}{}_\alpha
- \varepsilon^{\mu_1 \mu_3 \mu_5 \nu}
\varepsilon^{\mu_2 \mu_4 \mu_6 \mu}\right)
\varphi_{\mu_1} \varphi_{\mu_2}
\varphi_{\mu_3 \mu_4}\varphi_{\mu_5 \mu_6}
\nonumber
\\
&& - \varphi_{\mu_1} \left[
\varepsilon^{\mu_1 \mu_3 \mu_5 \alpha}
\varepsilon^{\mu_2 \nu \mu_6}{}_{\alpha}
\left( \varphi^\mu \varphi_{\mu_2}
\varphi_{\mu_5 \mu_6} \right)_{;\mu_3}
+
\varepsilon^{\mu_1 \mu_3 \mu_5 \alpha}
\varepsilon^{\mu_2 \mu \mu_6}{}_{\alpha}
\left( \varphi^\nu \varphi_{\mu_2}
\varphi_{\mu_5 \mu_6} \right)_{;\mu_3} \right]
\nonumber
\\
&&+ \varepsilon^{\mu_1 \mu \mu_5 \alpha}
\varepsilon^{\mu_2 \nu \mu_6}{}_{\alpha}
\left( \varphi^\sigma \varphi_{\mu_1}\varphi_{\mu_2}
\varphi_{\mu_5 \mu_6} \right)_{;\sigma}.
\label{D:Tmunu}
\end{eqnarray}
In the remainder of this subsection we simply denote $T^{\mu
\nu}_{(4,0)}$ by $T^{\mu \nu}$.

The first step is to determine explicitly the different order
time derivatives which appear in the zero-zero component of the
Einstein equation $G^{00} = T^{00}$, recalling that in $G^{00}$
there are at most first time derivatives of the metric. From
(\ref{D:Tmunu}) it follows that
\begin{eqnarray}
T^{00} &=&
\frac{1}{2} g^{00} \left(\varepsilon^{\mu_1 \mu_3 \mu_5 \alpha}
\varepsilon^{\mu_2 \mu_4 \mu_6}{}_\alpha \varphi_{\mu_1}
\varphi_{\mu_2} \varphi_{\mu_3 \mu_4}
\varphi_{\mu_5 \mu_6} \right)
- \varepsilon^{ijk0}\varepsilon^{pqr0}
\varphi_i \varphi_p \varphi_{jq} \varphi_{kr}
\nonumber\\
&& -2 \; \varphi_{\mu_1} \left[
\varepsilon^{\mu_1 \mu_3 \mu_5 k}
\varepsilon^{i0j}{}_k
(\varphi^0 \varphi_i \varphi_{\mu_5 j})_{;\mu_3} \right]
+\varepsilon^{i0jk}
\varepsilon^{p0r}{}_k(\varphi^\sigma
\varphi_i \varphi_p \varphi_{jr})_{;\sigma} \; ,
\label{D:T00}
\end{eqnarray}
{}from which we immediately see that $T^{00}$ contains no terms
in $\dddot{\varphi}$ nor in $\ddot{\varphi}_i$ (where latin
indices mean spatial components). This latter term could be
generated from the term within square brackets, but that would
require $\mu_3 = \mu_5=0$, in which case the result vanishes by
antisymmetry of the Levi-Civita tensor. Thus the highest order
time derivative of the scalar field it contains is
$\ddot{\varphi}$, whose coefficient can be determined directly
{}from (\ref{D:T00}) and will be given below. Regarding the
metric, there are second-order time derivatives coming from the
third-order covariant derivatives of $\varphi$ on the second
line, since $\varphi_{\alpha \beta \gamma} \supset -
(\partial_\gamma \Gamma^{\mu}_{\alpha \beta}) \varphi_\mu$. We
must therefore take $\mu_3=0$ and $\sigma=0$ to find these, and
we obtain
\begin{eqnarray}
\left. T^{00} \right| _{\ddot{g}_{ij}} &=&
\frac{\varphi^0}{N^2} \left[ \varepsilon^{pqk}
\varepsilon^{ij}{}_{k} \varphi_{i}
\varphi_p (\partial_0 \Gamma^{\nu}_{qj})\varphi_\nu\right] ,
\label{D:nasty00}
\end{eqnarray}
where $N\equiv1/\sqrt{-g^{00}}$ is the usual lapse in the ADM
decomposition, see Eq.~(\ref{ADM}), and $\varepsilon^{ijk}$ is
the 3-dimensional Levi-Civita tensor related to the 4-dimensional
one by
\begin{equation}
\varepsilon^{0ijk} = -\frac{\varepsilon^{ijk}}{N}.
\end{equation}
The subscript $\ddot{g}_{ij}$ on the left hand side of
(\ref{D:nasty00}) is due to the fact that the Christoffel symbols
$\Gamma^{\nu}_{qj}$ only contain first time derivatives of the
{\it spatial} components of the metric (see Appendix
\ref{AppA})\footnote{Hence note that $T^{00}$ contains no terms
in $\ddot{N}$ or $\ddot N_i$ (where $N$ and $N_i$ are the usual
lapse and shift in the ADM decomposition). Hence there are no
terms in $\ddot{g}_{00}$ nor $\ddot{g}_{0i}$.}. More explicitly
\begin{equation}
(\partial_0 \Gamma^{\nu}_{qj})\varphi_\nu = -N \varphi^0
(\partial_0 K_{qj}) + {\text{first-order derivatives}},
\end{equation}
where $K_{ij}$ is the extrinsic curvature. Thus
\begin{eqnarray}
\left. T^{00} \right| _{\ddot{g}_{ij}} &=&
-\frac{(\varphi^0)^2}{N} \left(
\varepsilon^{pqk} \varepsilon^{ij}{}_{k}
\varphi_{i} \varphi_p \dot{K}_{qj} \right)
\label{nasty00a}.
\end{eqnarray}

The second step involves carrying out the same procedure for
$\varphi_i T^{0i}$. From (\ref{D:Tmunu}) with $\mu=0$ and $\nu=i$
we find
\begin{eqnarray}
\varphi_iT^{0i} &=&
\frac{1}{2} \varphi_i g^{0i}
\left(\varepsilon^{\mu_1 \mu_3 \mu_5 \alpha}
\varepsilon^{\mu_2 \mu_4 \mu_6}{}_\alpha
\varphi_{\mu_1} \varphi_{\mu_2} \varphi_{\mu_3 \mu_4}
\varphi_{\mu_5 \mu_6} \right)
- \varepsilon^{\mu_1 \mu_3 \mu_5 i}
\varepsilon^{pqr0}\varphi_i \varphi_{\mu_1}
\varphi_p \varphi_{\mu_3 q} \varphi_{\mu_5 r}
\nonumber
\\
&&- \varphi_{\mu_1}\varphi_i \left[
\varepsilon^{\mu_1 \mu_3 \mu_5 \alpha}
\varepsilon^{\mu_2 i \mu_6}{}_\alpha
(\varphi^0 \varphi_{\mu_2}
\varphi_{\mu_5 \mu_6})_{;\mu_3} \right]
- \varphi_{\mu_1}\varphi_i \left[
\varepsilon^{\mu_1 \mu_3 \mu_5 k}
\varepsilon^{p 0 q}{}_k
(\varphi^i \varphi_{p}
\varphi_{\mu_5q})_{;\mu_3} \right]
\nonumber
\\
&&+\varphi_i \varepsilon^{p0qk}
\varepsilon^{\mu_2 i \mu_6}{}_k
(\varphi^\sigma \varphi_p \varphi_{\mu_2}
\varphi_{q \mu_6})_{;\sigma}.
\label{D:T0i}
\end{eqnarray}
Again, it is clear that there are no terms in $\dddot{\varphi}$.
Similarly there are no terms in $\ddot{\varphi}_j$ since one
always ends up with a contraction $\varepsilon^{ijk}\varphi_i
\varphi_j=0$. There are obviously terms in $\ddot{\varphi}$ (see
below). Concerning the terms in $\ddot{g}_{ij}$, following the
same logic as above, we find that they are given by
\begin{eqnarray}
\left. \varphi_i T^{0i} \right| _{\ddot{g}_{ij}} &=&
\frac{\varphi_m \varphi^m}{N^2} \left[ \varepsilon^{jrk}
\varepsilon^{pq}{}_{k}
\varphi_{j} \varphi_p (\partial_0 \Gamma^{\nu}_{rq})\varphi_\nu
\right].
\label{D:nasty0i}
\end{eqnarray}
It follows from (\ref{D:nasty00}) and (\ref{D:nasty0i}) that
$\varphi_i T^{0i} $ and $ T^{00}$ contain exactly the {\it same}
combination of second-order derivatives of the metric.

Furthermore, we recall that the components $G^{0\mu}$ of the
Einstein tensor do not involve second time derivatives of the
metric. Indeed, the Bianchi identities imply the covariant
conservation of the Einstein tensor, $\nabla_\lambda
G^{\lambda\mu} = 0$, therefore
\begin{equation}
\partial_0 G^{0\mu} =
- \partial_i G^{i\mu} - \mathcal{O}(\Gamma G),
\label{conserv}
\end{equation}
where $\mathcal{O}(\Gamma G)$ means the four terms involving
contractions of Christoffel symbols $\Gamma^\lambda_{\mu\nu}$
with the Einstein tensor $G^{\rho\sigma}$. Since the right-hand
side of Eq.~(\ref{conserv}) contains at most second time
derivatives of the metric tensor, this must be so for the
left-hand side $\partial_0 G^{0\mu}$, therefore the components
$G^{0\mu}$ contain at most first time derivatives.

Thus we arrive at the first important conclusion that, using
Eqs.~(\ref{D:nasty00}) and (\ref{D:nasty0i}), the combination of
the Einstein equations
\begin{equation}
\varphi^0 \varphi_i T^{0i}
- (\varphi_i \varphi^i) T^{00}
=\varphi^0 \varphi_\mu T^{0\mu}
- (\varphi_\mu \varphi^\mu) T^{00}
= \varphi^0 (\varphi_\mu G^{0\mu})
- (\varphi_\mu \varphi^\mu) G^{00}
\label{theoneperhaps}
\end{equation}
determines $\ddot{\varphi}$ in terms of {\it first} time
derivatives. As a result, any $\dddot{\varphi}$ appearing in the
equations of motion can be expressed in terms of second time
derivatives simply by differentiating (\ref{theoneperhaps}).

The next step is the determination of the coefficients of these
$\ddot{\varphi}$ terms. Starting from (\ref{D:T00}) and
(\ref{D:T0i}) we find
\begin{eqnarray}
\varphi_0 T^{00} &=&
\frac{\varphi_0 \varphi^0}{N^2} \left[
\varepsilon^{mnk} \varepsilon^{ij}{}_{k}
\varphi_{i} \varphi_m (\partial_0
\Gamma^{\nu}_{nj})\varphi_\nu\right]
+ B\varphi_0 (\varphi^0\ddot{\varphi})
+{\text{lower order derivatives}},
\label{D:h}
\\
\varphi_\mu T^{0\mu} &=&
\frac{\varphi_\lambda \varphi^\lambda}{N^2}
\left[ \varepsilon^{mnk} \varepsilon^{ij}{}_{k}
\varphi_{i} \varphi_m (\partial_0
\Gamma^{\nu}_{nj})\varphi_\nu\right]
- A (\varphi^0 \ddot{\varphi})
+ {\text{lower order derivatives}},
\label{D:f}
\end{eqnarray}
where
\begin{eqnarray}
B= - \frac{1}{N^2} \varepsilon^{mnk}
\varepsilon^{ij}{}_{k} \varphi_i
\varphi_m \Gamma^{0}_{nj} \, , \qquad
&&
A= - \frac{1}{N^2}\varepsilon^{mnk}
\varepsilon^{ij}{}_{k} \varphi_i \varphi_m \varphi_{nj} .
\label{Adef}
\end{eqnarray}
(Note that $A$ is in fact nothing other than ${\cal L}_{(3,0)}$
with only spatial indices.) Hence the second combination of
Einstein's equations
\begin{equation}
B (\varphi_\mu T^{0\mu}) + A T^{00}
= B (\varphi_\mu G^{0\mu}) + A G^{00}
\label{D:next}
\end{equation}
is an equation for the specific combination of $\ddot{g}_{ij}$
appearing in
\begin{equation}
\varepsilon^{mnk} \varepsilon^{ij}{}_{k}
\varphi_{i} \varphi_m (\partial_0 \Gamma^{\nu}_{nj})\varphi_\nu
= -N\varepsilon^{mnk} \varepsilon^{ij}{}_{k}
\varphi_{i} \varphi_m \varphi^0 \dot{K}_{nj},
\label{D:com}
\end{equation}
in terms of first time derivatives of the fields.

As a final step we must show that the third-order time
derivatives of the metric appearing in the equation of motion
${\cal E}=0$ for the scalar field, are exactly given by the time
derivative of the combination appearing in (\ref{D:com}). The
third-order derivatives of the metric in ${\cal E}$ are
\cite{Deffayet:2009mn}
\begin{equation}
{\cal E} \sim \varepsilon^{\mu_{\vphantom{()}1}
\mu_{\vphantom{()}3} \mu_{\vphantom{()}5} \alpha}
\varepsilon^{\mu_{\vphantom{()}2}
\mu_{\vphantom{()}4} \mu_{\vphantom{()}6}}{}_{\alpha}
\varphi_{\mu_{\vphantom{()}1}} \varphi_{\mu_{\vphantom{()}2}}
\varphi^\lambda R_{\mu_{\vphantom{()}3} \mu_{\vphantom{()}5}
\mu_{\vphantom{()}4} \mu_{\vphantom{()}6} ; \lambda},
\end{equation}
where $R_{\mu\nu\rho\sigma}$ is the Riemann tensor. To find the
$\dddot{g}_{ij}$ terms appearing here, it is sufficient to set
$\lambda=0$, and then the relevant third-order time derivative is
simply the derivative of the term in $\ddot{g}$ appearing in
\begin{equation}
\varepsilon^{\mu_1 \mu_3 \mu_5 \alpha}
\varepsilon_{\mu_2 \mu_4 \mu_6 \alpha}
\varphi_{\mu_1} \varphi_{\mu_2}
\varphi^0 R_{\mu_3 \mu_5 \mu_4 \mu_6}.
\end{equation}
However, using the results of Appendix \ref{AppA}, this is
nothing other than the combination (\ref{D:com}) (up to
irrelevant numerical factors). Since this is the contraction
appearing in (\ref{D:next}), it can thus be expressed in terms of
first time derivatives of the field. Thus the third-order time
derivatives of the metric appearing in the scalar field equation
of motion can be replaced by second-order time derivatives on
using the derivative of (\ref{D:next}).

Hence we arrive at the conclusion that, despite containing
higher-order time derivatives, all 11 equations of motion for
this theory can be expressed solely in terms of second-order time
derivatives: the derivative of the combination
(\ref{theoneperhaps}) gives $\dddot{\varphi}$ in terms of
second-order time derivatives, whilst the derivative of the
combination (\ref{D:next}) gives the required contraction of
$\dddot{g}_{ij}$ in terms of second-order time derivatives.

We now generalize these results to arbitrary $D$, $n$ and $p$,
which requires us to introduce more powerful notation. We will
also consider arbitrary linear combinations of these Lagrangians.

\subsection{General case: arbitrary $D$, $n$ and $p$}
\label{SubSec2B}

For the following discussion, the most convenient
expression~\cite{Deffayet:2009mn} for the Lagrangians ${\cal
L}_{(n+1,p)}$ uses the Levi-Civita fully antisymmetric tensor
\begin{equation}
\varepsilon^{\mu_{\vphantom{()}1} \mu_{\vphantom{()}2} \ldots
\mu_{\vphantom{()}D}} = - \frac{1}{\sqrt{-g}}
\delta^{[\mu_{\vphantom{()}1}}_1 \delta^{\mu_{\vphantom{()}2}}_2
\ldots \delta^{\mu_{\vphantom{()}D}]}_D,
\label{epsilon}
\end{equation}
where the square bracket denotes unnormalized permutations.
In any dimension $D\geq n$, we define
\begin{equation} \label{Lnp}
{\cal L}_{(n+1,p)} = -\mathcal{A}_{(2n)}
\varphi_{\mu_1} \varphi_{\mu_2} \mathcal{R}_{(p)}
\mathcal{S}_{(q)},
\end{equation}
where $\mathcal{A}_{(2n)}$ is a compact notation for
\begin{equation}
\mathcal{A}_{(2n)}^{\mu_{\vphantom{()}1} \mu_{\vphantom{()}2}
\ldots \mu_{\vphantom{()}2n}} \equiv
\frac{1}{(D-n)!}\,
\varepsilon^{\mu_{\vphantom{()}1}
\mu_{\vphantom{()}3} \mu_{\vphantom{()}5} \ldots
\mu_{\vphantom{()}2n-1}\, \nu_{\vphantom{()}1}
\nu_{\vphantom{()}2}\ldots
\nu_{\vphantom{()}D-n}}_{\vphantom{\nu_{\vphantom{()}1}}}
\,\varepsilon^{\mu_{\vphantom{()}2} \mu_{\vphantom{()}4}
\mu_{\vphantom{()}6} \ldots
\mu_{\vphantom{()}2n}}_{\hphantom{\mu_{\vphantom{()}2}
\mu_{\vphantom{()}4} \mu_{\vphantom{()}6} \ldots
\mu_{\vphantom{()}2n}}\nu_{\vphantom{()}1}
\nu_{\vphantom{()}2}\ldots \nu_{\vphantom{()}D-n}},
\label{calA}
\end{equation}
and where
\begin{eqnarray}
\mathcal{R}_{(p)} &\equiv&
\left(\varphi_{\lambda} \varphi^{\lambda}\right)^p\,
\prod_{i=1}^{p}
R_{\mu_{4i-1} \;\mu_{4i+1}\;\mu_{4i}\;\mu_{4i+2}},
\label{calR}\\
\mathcal{S}_{(q)} &\equiv& \prod_{i=0}^{q-1} \varphi_{\mu_{2n-1-2i}\;
\mu_{2n-2i}}
\label{calS},
\end{eqnarray}
with
\begin{equation}
q= n-1-2p.
\label{q}
\end{equation}
These definitions assume that $n$, $p$ and $q$
are integers in the ranges
\begin{equation}
1 \leq n \leq D, \quad 1 \leq p\leq
\left\lfloor\frac{n-1}{2}\right\rfloor, \quad \text{and}
\quad 1 \leq q\leq n-1.
\end{equation}
We also set $\mathcal{R}_{(0)} \equiv 1$ for $p=0$ and
$\mathcal{S}_{(0)} \equiv 1$ for $q=0$, and use the convention
that $\mathcal{R}_{(p)} = 0$ for $p < 0$ and $\mathcal{S}_{(q)} =
0$ for $q < 0$. [The cases $n = 0$ and $n = -1$, with $p= 0$, may
also be defined~\cite{Nicolis:2008in} as ${\cal L}_{(1,0)} =
\varphi$ and ${\cal L}_{(0,0)} = \text{const.}$, but we do not
consider them here since they obviously do not yield higher-order
field equations.] The numerical factor $1/(D-n)!$ is introduced
in Eq.~(\ref{calA}) so that $\mathcal{A}_{(2n)}$ keeps the same
expression in terms of products of metric tensors in any
dimension.

When $p = 0$, i.e., without any Riemann tensor involved, these
definitions reduce to the Galileons of
Ref.~\cite{Nicolis:2008in}. For instance, ${\cal
L}_{(2,0)}=\varphi_\mu \varphi^\mu$ is the kinetic term of a
standard scalar field (though, referring to Eq.~(\ref{SGal}), one
should choose a negative $k_{(2,0)}$ in order not to have a
ghost around an empty and flat background). The cubic Lagrangian
${\cal L}_{(3,0)}= \varphi_\mu \varphi^\mu\Box \varphi -
\varphi_\mu \varphi^{\mu\nu}\varphi_\nu = \frac{3}{2} \varphi_\mu
\varphi^\mu\Box \varphi+ \text{tot. div.}$, is the one obtained
in the decoupling limit of the DGP model
\cite{Dvali:2000hr,Luty:2003vm,Nicolis:2004qq}. The quartic
Lagrangian was given in Eq.~(\ref{D:L4}). In $D= 4$ dimensions,
there also exists ${\cal L}_{(5,0)}$, written for instance in
Eq.~(3) of \cite{Deffayet:2009mn}. On the other hand, the
Lagrangians (\ref{Lnp}) with $p \neq 0$, involving one or several
Riemann tensors, are the ``counterterms'' introduced in
Refs.~\cite{Deffayet:2009wt,Deffayet:2009mn,Deffayet:2010zh,
Deffayet:2011gz,Deffayet:2013lga} to avoid any third derivative
in the field equations. There are only two of them in $D=4$
dimensions, ${\cal L}_{(4,1)}$ and ${\cal L}_{(5,1)}$,
cf.~Eqs.~(14) and (15) of \cite{Deffayet:2009mn}. Our proof below
will be valid for all ${\cal L}_{(n+1,p)}$ in any dimension $D$,
as well as linear combinations of them.

\subsubsection{ ${\cal L}_{(n+1,p)}$ and their linear combinations}

We first focus on a single Lagrangian ${\cal L}_{(n+1,p)}$.
In order to write its Einstein equations $G^{\alpha\beta} =
T^{\alpha\beta}$ in the simplest way, it will be useful to
introduce the following compact notation generalizing
(\ref{calA}):
\begin{eqnarray}
\mathcal{A}^{\alpha}_{(2n,i)} &\equiv&
\mathcal{A}_{(2n)}^{\mu_{\vphantom{()}1} \mu_{\vphantom{()}2}
\ldots \mu_{\vphantom{()}i-1}\; \alpha\;
\mu_{\vphantom{()}i+1} \ldots \mu_{\vphantom{()}2n}},
\label{Ai}\\
\mathcal{A}^{\alpha\beta}_{(2n,i,j)} &\equiv&
\mathcal{A}_{(2n)}^{\mu_{\vphantom{()}1} \mu_{\vphantom{()}2}
\ldots \mu_{\vphantom{()}i-1}\; \alpha\; \mu_{\vphantom{()}i+1}
\ldots \mu_{\vphantom{()}j-1}\; \beta\;
\mu_{\vphantom{()}j+1} \ldots \mu_{\vphantom{()}2n}},
\label{Aij}
\end{eqnarray}
where $2n$ is the rank of the tensor, and $i$ and $j$ locate the
positions of the indices $\alpha$ and $\beta$ which are
explicitly indicated. The energy-momentum tensor is then given by
\begin{eqnarray}
T^{\alpha\beta} &=&\left[\left(\frac{1}{2}g^{\alpha\beta}
+p\, \frac{\varphi^\alpha\varphi^\beta}{\varphi_\lambda^2}\right)
\mathcal{A}_{(2n)}
-\mathcal{A}_{(2n+2, 2n+1, 2n+2)}^{\alpha\beta}\right]
\varphi_{\mu_1} \varphi_{\mu_2} \mathcal{R}_{(p)}
\mathcal{S}_{(q)}\nonumber\\
&&- q\, \mathcal{A}_{(2n, 4p+4)}^{(\alpha}\varphi_{\mu_1}
\left[\varphi^{\beta)} \varphi_{\mu_2} \mathcal{R}_{(p)}
\mathcal{S}_{(q-1)}\right]_{;\mu_{4p+3}}\nonumber\\
&&+\frac{q}{2}\mathcal{A}^{\alpha\beta}_{(2n,4p+3,4p+4)}
\left[\varphi^\sigma \varphi_{\mu_1} \varphi_{\mu_2} \mathcal{R}_{(p)}
\mathcal{S}_{(q-1)}\right]_{;\sigma}\nonumber\\
&&+2p\,\mathcal{A}^{(\alpha\beta)}_{(2n,4p+1,4p+2)}
\left[\varphi_{\mu_1} \varphi_{\mu_2}
(\varphi_\lambda^2)\mathcal{R}_{(p-1)}
\mathcal{S}_{(q)}\right]_{;\mu_{4p}\;\mu_{4p-1}}\nonumber\\
&&-p\,\mathcal{A}^{(\alpha}_{(2n,4p-1)}
R^{\beta)}_{\hphantom{\beta)}\mu_{4p+1}\;\mu_{4p}\;\mu_{4p+2}}\,
\varphi_{\mu_1} \varphi_{\mu_2}
(\varphi_\lambda^2) \mathcal{R}_{(p-1)} \mathcal{S}_{(q)},
\label{Tmunu}
\end{eqnarray}
where symmetrization over $\alpha$ and $\beta$ is assumed [i.e.,
$X^{(\alpha\beta)} = (X^{\alpha\beta}+X^{\beta\alpha})/2$ for any
tensor], notably on the second and last two lines which are not
automatically symmetric. Regarding the first term, note that the
factor $p/\varphi_\lambda^2$ does not cause any divergence: it
vanishes when $p=0$, and when $p\neq0$ it is multiplied by
$(\varphi_\lambda^2)^p$ contained in the $\mathcal{R}_{(p)}$ term
(\ref{calR}). The tensor $\mathcal{A}_{(2n+2, 2n+1,
2n+2)}^{\alpha\beta}$ on the first line has $2n+2$ free indices,
the last two of them being $\alpha$ and $\beta$. The
corresponding term in $T^{\alpha \beta}$ originates from the
variation with respect to the metric of the contracted indices in
Eq.~(\ref{calA}). It vanishes if $D < n+1$. Note that in the
second line, the factor $\varphi_{\mu_1}$ has been extracted from
the square bracket which is covariantly derived with respect to
$\mu_{4p+3}$, because $\varphi_{\mu_1\;\mu_{4p+3}}$ with two odd
indices would vanish when contracted with the first antisymmetric
$\varepsilon$ tensor of Eq.~(\ref{calA}).

This energy-momentum tensor in Eq.~(\ref{Tmunu}) allows us to
prove several important lemmas. Let us focus to start with on
third derivatives: by suitably permuting and relabeling dummy
indices, one may rewrite them as
\begin{eqnarray}
T^{\alpha\beta}_\text{3rd der.} &=& \frac{q(q-1)}{2}
\mathcal{A}_{(2n,4p+3,4p+4)}^{\alpha\beta}
\varphi_{\mu_1} \varphi_{\mu_2}
\mathcal{R}_{(p)} \mathcal{S}_{(q-2)}
\varphi^{\lambda}\varphi_{\mu_{4p+5}\, \mu_{4p+6}\, \lambda}
\nonumber\\
&&+4 p^2 \mathcal{A}_{(2n, 4p+1,4p+2)}^{\alpha\beta}
\varphi_{\mu_1} \varphi_{\mu_2}
\mathcal{R}_{(p-1)} \mathcal{S}_{(q)}
\varphi^{\lambda} \varphi_{\lambda\,\mu_{4p}\, \mu_{4p-1}}.
\label{ThirdDer}
\end{eqnarray}
This is is equivalent to Eq.~(34) of Ref.~\cite{Deffayet:2009mn}.
We can thus arrive at the following conclusions.

First, no third-order derivative of the metric tensor enters
$T^{\alpha\beta}$. Indeed, the Bianchi identities
$R_{\lambda\mu[\nu\rho;\sigma]} = 0$ cancel most of the
differentiated Riemann tensors in (\ref{Tmunu}), because three of
their indices are contracted with the same antisymmetric
$\varepsilon$ tensor of $\mathcal{A}_{(2n)}$. The only non
vanishing ones come from the $\mathcal{R}_{(p);\sigma}$ of the
third term, but they exactly cancel with the derivatives of the
Riemann tensors generated by permuting the indices of
$\mathcal{S}_{(q);\mu_{4p}\;\mu_{4p-1}}$ coming from the fourth
term.

Second, Eq.~(\ref{ThirdDer}) also proves that $T^{00}$ does not
contain any $\dddot\varphi$ nor $\ddot\varphi_{i}$. Indeed,
because $\alpha = \beta = 0$, one odd index ($4p+3$ or $4p+1$)
and one even index ($4p+4$ or $4p+2$) must be~$0$, therefore all
other (contracted) indices of the two $\varepsilon$ tensors in
$\mathcal{A}_{(2n)}$ must be spatial. In conclusion, the only
third-differentiated scalar fields (\ref{ThirdDer}) cannot
contain more than one time derivative in $T^{00}$.

Third, the contraction $\varphi_\alpha T^{\alpha\beta}$ (and in
particular $\varphi_\alpha T^{\alpha0}$, that we will use in the
argument below) does not contain any third derivative of the
scalar field. Indeed, $\varphi_\alpha$ and $\varphi_{\mu_1}$ are
then contracted with the same antisymmetric $\varepsilon$ tensor
entering $\mathcal{A}_{(2n)}$, in Eq.~(\ref{ThirdDer}), making it
vanish.

We now proceed as in Section \ref{SubSec2A} and compute the
second time derivatives entering $\varphi_\alpha T^{\alpha 0}$
and $T^{00}$, using the full expression (\ref{Tmunu}). However,
before doing so, and in order to simplify the resulting
expressions, we introduce the following further notation. Using
(\ref{Aij}), we define
\begin{equation} \label{Lspatial}
{\cal L}_{(n+1,p)}^\text{spatial} \equiv
-\mathcal{A}_{(2n+2,2n+1,2n+2)}^{00}
\varphi_{\mu_1} \varphi_{\mu_2} \mathcal{R}_{(p)}
\mathcal{S}_{(q)},
\end{equation}
which is the analogue of Eq.~(\ref{Lnp}), but where now all
contracted indices are spatial (though the covariant derivatives
and the Riemann tensors remain $D$-dimensional). As before, see
Eq.~(\ref{q}), $q= n-1-2p$. Note that
$\mathcal{A}_{(2n+2,2n+1,2n+2)}^{00}$ has two extra indices
relative to $\mathcal{A}_{(2n)}$, namely the last two which are
both $0$, meaning that the $2n$ first must be spatial. Similarly,
we define
\begin{equation} \label{LGamma}
{\cal L}_{(n+1,p)}^\Gamma \equiv -\mathcal{A}_{(2n+4,2n+3,2n+4)}^{00}
\varphi_{\mu_1} \varphi_{\mu_2} \mathcal{R}_{(p)} \mathcal{S}_{(q)}
\Gamma^0_{\mu_{2n+1}\; \mu_{2n+2}},
\end{equation}
which, as before, contains $n+1$ scalar fields and $p$ Riemann
tensors, as well as now a Christoffel symbol $\Gamma^0_{ij}$.
Notice that all contracted indices again are spatial, because the
$(2n+3)$rd and $(2n+4)$th indices are imposed to be $0$.
[Actually, ${\cal L}_{(n+2,p)}^\text{spatial}$ contains
\hbox{$-(n-2p) \dot\varphi {\cal L}_{(n+1,p)}^\Gamma$}.]

On using (\ref{Tmunu}), the second time derivatives entering
$\varphi_\alpha T^{\alpha 0}$ and $T^{00}$ can then be written as
\begin{eqnarray}
\varphi_\alpha T^{\alpha 0}_\text{2nd t der.} &=&
-\varphi^0 \left(A\,\ddot\varphi
+\varphi_\alpha^2\, C\right),
\label{Talpha0}\\
T^{00}_\text{2nd t der.} &=& \varphi^0 \left(B\, \ddot\varphi
-\varphi^0\, C\right),
\label{T00}
\end{eqnarray}
where
\begin{eqnarray}
A &\equiv& \frac{\bar q(\bar q-1)}{2}
{\cal L}_{(n,p)}^\text{spatial}
+4p^2 {\cal L}_{(n,p-1)}^\text{spatial},
\label{G:A}\\
B &\equiv& \frac{\bar q(\bar q-1)}{2}
{\cal L}_{(n-1,p)}^\Gamma
+4p^2 {\cal L}_{(n-1,p-1)}^\Gamma,
\label{G:B}\\
C &\equiv&\frac{\bar q(\bar q-1)}{4}\,
\mathcal{A}_{(2n,4p+3,4p+4)}^{00}
\varphi_{\mu_1} \varphi_{\mu_2}
\mathcal{R}_{(p)} \mathcal{S}_{(\bar q-2)}
\ddot g_{\mu_{4p+5}\, \mu_{4p+6}}\nonumber\\
&&+2 p^2 \mathcal{A}_{(2n,4p-1,4p)}^{00}
\varphi_{\mu_1} \varphi_{\mu_2}
\mathcal{R}_{(p-1)} \mathcal{S}_{(\bar q)}
\ddot g_{\mu_{4p+1}\, \mu_{4p+2}},
\label{G:C}
\end{eqnarray}
with $\bar q \equiv n-1-2p$ in these expressions ---~but the
$q$'s involved within ${\cal L}^\text{spatial}$ and ${\cal
L}^\Gamma$ depend on their precise indices $(n,p)$ as defined in
Eqs.~(\ref{Lspatial}) and (\ref{LGamma}). Note that $A$ and $B$
involve at most first time derivatives, whereas $C$ contains a
specific contraction of $\ddot g_{ij}$ with other fields
(themselves differentiated at most once with respect to time).

The fact that the \textit{same} contraction $C$ of $\ddot g_{ij}$
enters both (\ref{Talpha0}) and (\ref{T00}) will be crucial
below. Whilst this looks miraculous, it is actually simply a
consequence of the diffeomorphism invariance of the theory,
together with the fact that $T^{\alpha\beta}$ involves at most
second derivatives of the metric as proved in
Eq.~(\ref{ThirdDer}). Indeed, diffeomorphism invariance implies
Eq.~(\ref{conservTmunu}), which tells us that
$\varphi^{\alpha}{\cal E} = 2\partial_0 T^{0 \alpha}+2\partial_i
T^{i \alpha} +\mathcal{O}\left(\Gamma T\right)$. The third time
derivatives of the metric entering $\mathcal{E}$ can thus come
only from $2\partial_0 T^{0 \alpha}$, proving that all $T^{0
\alpha}$ must contain exactly the same terms in $\ddot g_{ij}$,
multiplied by $\varphi^{\alpha}$. Using definition (\ref{G:C}),
we have explicitly $\left. T^{0\alpha} \right|_{\ddot{g}_{ij}} =
-\varphi^0\varphi^\alpha C$, while $\left. \mathcal{E}
\right|_{\dddot{g}_{ij}} =-2\varphi^0
\left.(\partial_0 C)\right|_{\dddot{g}_{ij}} $.

This dependence of (\ref{Talpha0}) and (\ref{T00})
on the same contraction $C$ of $\ddot g_{ij}$ allows us to
cancel it
in the linear combination
\begin{equation}
\varphi^0 \varphi_\alpha T^{\alpha 0} - \varphi_\alpha^2 T^{00}.
\label{comblin}
\end{equation}
On the other hand, this combination does contain a term
\begin{equation}
-\varphi^0\left(\varphi^0 A+\varphi_\alpha^2 B\right) \ddot\varphi
\label{ddotphi}
\end{equation}
proportional to the second time derivative of the scalar field.
Our lemmas above also prove that all its other terms involve at
most first time derivatives of the fields (and up to three
spatial derivatives, but they do not pose any difficulty for the
Cauchy problem). Now, using the fact that the $G^{\alpha0}$
components of the Einstein tensor depend only on first time
derivatives of the metric, cf.~Eq.~(\ref{conserv}), we arrive at
the following crucial combination of Einstein's equations:
\begin{equation}
\varphi_\alpha^2 (G^{00}-T^{00})
-\varphi^0 \varphi_\alpha (G^{\alpha 0}-T^{\alpha 0}) = 0.
\label{comblin2}
\end{equation}
This generalizes Eq.~(\ref{theoneperhaps}) for arbitrary $n$
and $p$, and allows us to express $\ddot\varphi$ in terms of
undifferentiated fields, their spatial derivatives, and their
\textit{first} time derivatives.

We now follow the same logic as in subsection \ref{SubSec2A}.
On taking the time derivative of Eq.~(\ref{comblin2}), we have a
way to express $\dddot\varphi$ in terms of fields which are
differentiated at most twice with respect to time. Since
$\dddot\varphi$ was the only higher-order time derivative
entering Einstein's equations $G^{\alpha\beta} =
T^{\alpha\beta}$, we conclude that all of them now become of
second order (as far as time is concerned). Moreover, since the
linear combination (\ref{comblin2}) does not depend on the values
$n$ and $p$ specifying the model ${\cal L}_{(n+1,p)}$, we can
repeat the same argument for any linear combination $\sum_{n,p}
k_{(n,p)} {\cal L}_{(n,p)}$, and this achieves our proof: Only
the fields and their first time derivatives should need to be
specified on an initial value surface.

Of course, our argument fails if the coefficient (\ref{ddotphi})
happens to vanish at some spacetime point, since $\ddot\varphi$
can no longer be expressed in terms of at most first time
derivatives of fields, in such a case. This is notably what
happens on a surface where $\varphi_{i} = 0$, which corresponds
precisely to the unitary gauge chosen in
Refs.~\cite{Gleyzes:2014dya,Gleyzes:2014qga,Lin:2014jga}.
However, no third time derivative enters any field equation when
$\varphi_{i} = 0$, therefore one can keep all of them without any
difficulty when this happens. Let us anyway mention that our
argument is \textit{generic}, as required when discussing the
well-posedness of the Cauchy problem. The spacetime domains where
Eq.~(\ref{ddotphi}) may vanish have measure zero, and our proof
is thus valid almost everywhere, in the mathematical sense of
measure theory.

Let us now show that the scalar field equation $\mathcal{E} = 0$
may also be recast as a second-order differential equation with
respect to time. We already saw in Eq.~(\ref{conservTmunu}) that
$\mathcal{E}$ can be obtained by taking the divergence of
Einstein's equations. However, we now show how the third time
derivatives of the metric tensor can be removed from
$\mathcal{E}$, while still keeping a non-zero term in
$\ddot\varphi$. For this purpose, consider another linear
combination of Einstein's equations
\begin{equation}
A(G^{00}-T^{00}) +B\, \varphi_\alpha(G^{\alpha 0}-T^{\alpha 0}) = 0,
\label{ddotgij}
\end{equation}
which generalizes Eq.~(\ref{D:next}) above for arbitrary $n$ and
$p$. All second time derivatives of the scalar field cancel in
this combination, but it still contains a term
\begin{equation}
\varphi^0\left(\varphi^0 A +\varphi_\alpha^2 B\right)C,
\label{ddotg}
\end{equation}
where $C$ involves a specific contraction of $\ddot g_{ij}$,
cf.~Eq.~(\ref{G:C}). Therefore, the linear combination
(\ref{ddotgij}) allows us to express this contraction in terms of
fields differentiated at most once with respect to
time.\footnote{Note that the same coefficient
$\varphi^0\left(\varphi^0 A +\varphi_\alpha^2 B\right)$ enters
both Eqs.~(\ref{ddotphi}) and (\ref{ddotg}). At the generic
spacetime points where it does not vanish, the two linear
combinations of Einstein's equations (\ref{comblin2}) and
(\ref{ddotgij}) can thus both be used, to extract $\ddot\varphi$
and the specific contraction $C$ of $\ddot g_{ij}$ in terms of
fields differentiated at most once with respect to time.} Or, by
taking the time derivative of (\ref{ddotgij}), one may express
the contraction (\ref{G:C}), with $\ddot g_{ij}$ replaced by
$\dddot g_{ij}$, in terms of at most second time derivatives.
Returning to the scalar field equation $\mathcal{E} = 0$, it
turns out that the third time derivatives entering it take
precisely the same form, namely:
\begin{eqnarray}
\mathcal{E}_\text{3rd t der.}&=&-\frac{1}{2}\varphi^0 \Bigl[q(q-1)
\mathcal{A}_{(2n,4p+3,4p+4)}^{00}
\varphi_{\mu_1} \varphi_{\mu_2} \mathcal{R}_{(p)} \mathcal{S}_{(q-2)}
\dddot g_{\mu_{4p+5}\, \mu_{4p+6}}\nonumber\\
&&\hphantom{-\frac{1}{2}\varphi^0 \Bigl[}+ 8 p^2
\mathcal{A}_{(2n,4p-1,4p)}^{00}
\varphi_{\mu_1} \varphi_{\mu_2} \mathcal{R}_{(p-1)} \mathcal{S}_{(q)}
\dddot g_{\mu_{4p+1}\, \mu_{4p+2}}\Bigr].
\label{dddotg}
\end{eqnarray}
As discussed below Eq.~(\ref{G:C}), this is a consequence
of the diffeomorphism invariance of the theory (and of the absence
of third time derivatives of the metric in $T^{\alpha\beta}$).
Therefore, all third time derivatives entering $\mathcal{E} = 0$
are exactly canceled by adding to it the time derivative of
Eq.~(\ref{ddotgij}) multiplied by $2/(\varphi^0 A +
\varphi_\alpha^2 B)$.

This procedure can also be extended to an arbitrary sum of
Lagrangians $\sum_{n,p} k_{(n,p)} {\cal L}_{(n,p)}$, although
this is less obvious than for the Einstein equations, cf.~the
paragraph below Eq.~(\ref{comblin2}). Indeed, since the
coefficients $A$ and $B$ of Eqs.~(\ref{G:A}) and (\ref{G:B})
depend on $n$ and $p$, the linear combination (\ref{ddotgij}) is
thus specific to a single case. However, if we denote as
$A_{(n,p)}$, $B_{(n,p)}$ and $C_{(n,p)}$ the coefficients
(\ref{G:A})--(\ref{G:C}) corresponding to a given Lagrangian
${\cal L}_{(n,p)}$, the above results immediately show that the
linear combination
\begin{equation}
\left(\sum_{n,p} k_{(n,p)} A_{(n,p)}\right) (G^{00}-T^{00})
+ \left(\sum_{n,p} k_{(n,p)} B_{(n,p)}\right)
\varphi_\alpha (G^{\alpha 0}-T^{\alpha 0}) = 0
\label{sumddotgij}
\end{equation}
does not contain any $\ddot\varphi$, whereas its terms involving
$\ddot g_{ij}$ are of the form
\begin{equation}
\varphi^0\left[\varphi^0
\left(\sum_{n,p} k_{(n,p)} A_{(n,p)}\right)
+ \varphi_\alpha^2
\left(\sum_{n,p} k_{(n,p)} B_{(n,p)}\right)\right]
\left(\sum_{n,p} k_{(n,p)} C_{(n,p)}\right).
\label{sumddotg}
\end{equation}
Moreover, all its other terms involve at most first time
derivatives. On the other hand, the third time derivatives
entering the scalar field equation $\mathcal{E} = 0$ are exactly
the same as those of $-2\varphi^0\sum_{n,p} k_{(n,p)} \partial_0
C_{(n,p)}$, again for the same reasons as explained below
Eq.~(\ref{G:C}). Therefore, it suffices to add to this scalar
field equation the time derivative of Eq.~(\ref{sumddotgij}),
multiplied by $2$ and divided by the large coefficient within the
square brackets of (\ref{sumddotg}), to get a second-order
differential equation with respect to time.

Together with our previous proof that the Einstein equations
themselves can be recast in terms of at most second time
derivatives, we thus arrive at the following powerful result:
\renewcommand{\labelitemi}{}
\begin{itemize}
\item All generalized
Galileon models $\sum_{n,p} k_{(n,p)} {\cal L}_{(n,p)}$ in curved
spacetime, with arbitrary constant coefficients $k_{(n,p)}$,
yield $\frac{D(D+1)}{2}+1$ field equations which can be combined
so that none of them involve more than second time derivatives.
\end{itemize}
Obviously, this does not prove that all these theories are
stable. For instance, it suffices to choose the wrong signs for
some of the coefficients $k_{(n,p)}$ to get a ghost scalar degree
of freedom, notably in the simplest case $+{\cal L}_{(2,0)}$ with
a positive sign. But this shows that the specific combination
$\sum_{p=0}^{p_{max}} \mathcal{C}_{(n,p)} {\cal L}_{(n,p)}$ of
Ref.~\cite{Deffayet:2010zh} is not safer, nor worse, than any
other linear combination. The only difference is that the field
equations of these ``covariantized'' Galileons (\ref{CovG}) do
not involve \textit{any} third derivative, even purely spatial,
whereas we proved that arbitrary sums $\sum_{n,p} k_{(n,p)} {\cal
L}_{(n,p)}$ can be cured from their third \textit{time}
derivatives, but they keep other types of third derivatives
(spatial, or mixing space and time).

\subsubsection{Horndeski/k-essence-like generalizations}

It was shown in Ref.~\cite{Deffayet:2011gz} (see also the review
\cite{Deffayet:2013lga}), that the above models can be
generalized further. Indeed, the flat-space Lagrangians ${\cal
L}_{(n+1,0)}$, with $p = 0$, can first be integrated by parts to
be rewritten as $-\frac{n+1}{2}\,\varphi_\lambda^2\,
\mathcal{A}_{(2n-2)}^{\mu_{\vphantom{()}3} \mu_{\vphantom{()}4}
\ldots \mu_{\vphantom{()}2n}} \mathcal{S}_{(n-1)} + \text{tot.
div.}$, and one may then replace the factor $\varphi_\lambda^2$
by any function $f(\varphi, \varphi_\lambda^2)$ without changing
the structure of the higher derivatives of the model. Hence the
simplest case of ${\cal L}_{(2,0)}=\varphi_\lambda^2$ generates
all k-essence theories $f(\varphi, \varphi_\lambda^2)$, which
also include the tadpole ${\cal L}_{(1,0)}=\varphi$ and the
cosmological constant ${\cal L}_{(0,0)}=\text{const}$.
The $p\neq 0$ cases may also be generalized in the
same way~\cite{Deffayet:2011gz}, as
\begin{equation}
-f_{(n+1,p)}
(\varphi, \varphi_\lambda^2)\,
\mathcal{A}_{(2n-2)}^{\mu_{\vphantom{()}3} \mu_{\vphantom{()}4}
\ldots \mu_{\vphantom{()}2n}} \mathcal{R}_{(p)} \mathcal{S}_{(q)}.
\label{Horndeski}
\end{equation}
In $D = 4$ dimensions, one recovers the full class of Horndeski's
theories~\cite{Horndeski}, where the functions $f_{(n+1,p)}$ of
$\varphi$ and $\varphi_\lambda^2$ multiplying the different terms
need to have specific relations amongst themselves in order to
avoid the appearance of third derivatives in the field equations.
Another possible generalization of the $p\neq 0$ cases has been
considered in Refs.~\cite{Gleyzes:2014dya,Gleyzes:2014qga}, and
is obtained by the following set of Lagrangians\footnote{Note
that the set of theories defined respectively by Lagrangians
(\ref{Horndeski}) and (\ref{genGalbis}) can be related to each
other by suitable integration by parts and identities between
Levi-Civita tensors of the kind discussed in~\cite{Deffayet:2011gz}.}
\begin{equation}
f_{(n+1,p)}(\varphi, \varphi_\lambda^2)\,
{\cal L}_{(n+1,p)} = -f_{(n+1,p)}(\varphi, \varphi_\lambda^2)\,
\mathcal{A}_{(2n)} \varphi_{\mu_1} \varphi_{\mu_2}
\mathcal{R}_{(p)} \mathcal{S}_{(q)},
\label{genGalbis}
\end{equation}
which are just obtained by multiplying the Lagrangians of
Eq.~(\ref{Lnp}) by arbitrary functions $f_{(n+1,p)}$.
The novelty claimed by
Refs.~\cite{Gleyzes:2014dya,Gleyzes:2014qga,Lin:2014jga} is that
even without any relation between such functions, these theories
in any case do not generate any second scalar degree of freedom,
although their field equations do involve third derivatives, and
notably third time derivatives. These references concluded that
the most general model in $D=4$ dimensions depends on six
arbitrary functions of $\varphi$ and $\varphi_\lambda^2$.
However, as mentioned in our Introduction, the arguments given in
Refs.~\cite{Gleyzes:2014dya,Gleyzes:2014qga,Lin:2014jga} do not
appear to be fully convincing.

Our reasoning above can be repeated without much change for these
wide classes of theories. Indeed, let us first consider a single
Lagrangian of the form (\ref{genGalbis}). It is easy to see that
the energy-momentum tensor of such a theory is just obtained by
Eq.~(\ref{Tmunu}) where one replaces everywhere $\varphi_{\mu_2}$
by $f_{(n+1,p)}\varphi_{\mu_2}$, and to which one adds the extra
term
\begin{equation}
-\frac{\partial f}{\partial X}\,
\varphi^\alpha \varphi^\beta \mathcal{L}_{(n+1,p)},
\end{equation}
where $X\equiv \varphi_\lambda^2$, and $\mathcal{L}_{(n+1,p)}$ is
the generalized Galileon Lagrangian~(\ref{Lnp}) (corresponding to
$f = 1$). The three lemmas derived below Eq.~(\ref{ThirdDer})
follow then immediately: No third derivative of the metric enters
$T^{\alpha\beta}$, and neither $T^{00}$ nor $\varphi_\alpha
T^{\alpha 0}$ contain\footnote{On the other hand, there do remain
some $\ddot\varphi_i$ in $\varphi_\alpha T^{\alpha 0}$ when
considering a Lagrangian of the form (\ref{Horndeski}). However,
even in this case, our reasoning also shows that the linear
combination (\ref{comblin2}) does not contain any second time
derivative but those of the scalar field.} any $\dddot\varphi$
or $\ddot\varphi_i$. For the same reasons as explained below
Eq.~(\ref{G:C}), one also finds that all components $T^{\alpha
0}$ of the stress-energy tensor involve the same contraction of
$\ddot g_{ij}$, namely
\begin{equation}
\left. T^{\alpha 0} \right| _{\ddot{g}_{ij}}
= - \varphi^\alpha \varphi^0
\left(f\, C + \frac{\partial f}{\partial X}
\left.\mathcal{L}_{(n+1,p)}
\right| _{\ddot{g}_{ij}} \right),
\label{Ta0Horndeski}
\end{equation}
where $C$ is given in Eq.~(\ref{G:C}). Therefore, these second
time derivatives of the metric still cancel in the same linear
combination (\ref{comblin2}) as above, while it is easy to check
that the coefficient of $\ddot\varphi$ remains generically non
zero (although it is changed with respect to the $f=1$ case). In
conclusion, Eq.~(\ref{comblin2}) still allows us to express
$\ddot\varphi$ in terms of undifferentiated fields, their spatial
derivatives, and their \textit{first} time derivatives, and the
time derivative of this Eq.~(\ref{comblin2}) can thus be used to
remove all third time derivatives entering Einstein's equations.
Moreover, since the linear combination (\ref{comblin2}) does not
depend on the precise model (\ref{genGalbis}) under
consideration, one can obviously repeat the same argument
for any sum of such Lagrangians.

Let us now consider the scalar field equation $\mathcal{E} = 0$
deriving from a Lagrangian (\ref{genGalbis}). It involves third
time derivatives of the metric, of the form
\begin{equation}
\left. \mathcal{E} \right| _{\dddot{g}_{ij}} = - 2 \varphi^0
\left.\partial_0\left(f\, C + \frac{\partial f}{\partial X}
\mathcal{L}_{(n+1,p)}\right)\right| _{\dddot{g}_{ij}}.
\end{equation}
They are proportional to those entering $\partial_0 T^{\alpha
0}$, Eq.~(\ref{Ta0Horndeski}), and this is not a surprise since
this is again a consequence of diffeomorphism invariance, as
explained below Eq.~(\ref{G:C}). It thus suffices to construct
the unique linear combination of the Einstein equations
$(G^{00}-T^{00}) = 0$ and $\varphi_\alpha(G^{\alpha 0} -T^{\alpha
0}) = 0$ such that all $\ddot\varphi$ cancel, and the time
derivative of this combination allows us to replace all third
derivatives entering $\mathcal{E} = 0$ in terms of at most second
time derivatives. This conclusion can also be extended to an
arbitrary sum of Lagrangians (\ref{genGalbis}), by following the
same reasoning as in Eqs.~(\ref{sumddotgij})--(\ref{sumddotg})
above. [Let us also recall that for any theory, the scalar field
equation $\mathcal{E} = 0$ can always be recovered from the
divergence of Einstein's equations, Eq.~(\ref{conservTmunu}).]

In conclusion, any sum of Lagrangians $\sum_{n,p}
f_{(n,p)}\left(\varphi,\varphi_\lambda^2\right) {\cal
L}_{(n,p)}$, i.e., the k-essence-like extensions of generalized
Galileon theories (\ref{Lnp}), yield field equations which can be
recast as a set of $\frac{D(D+1)}{2}+1$ second-order differential
equations (as far as time is concerned). In $D$ dimensions, there
are $\lfloor\frac{D+1}{2}\rfloor\lfloor\frac{D}{2}+1\rfloor$
classes of models, each depending on an arbitrary function
$f_{(n,p)}(\varphi, \varphi_\lambda^2)$. This is
consistent with the 6 classes found in
Refs.~\cite{Gleyzes:2014dya,Gleyzes:2014qga,Lin:2014jga} for
$D = 4$, and would give for instance 30 classes of models in
the $D=10$ dimensions of string theory.

\section{Hamiltonian analysis of the quartic Galileon}
\label{Sec3}

In this final section, we work in $D=4$ dimensions and focus
solely on the Lagrangian ${\cal L}_{(4,0)}$. Our aim is to carry
out a Hamiltonian analysis of this particular theory without
fixing any gauge, and to show ---~from a Hamiltonian point of
view~--- that it {\it cannot} contain 4 degrees of freedom. As
such, the results of this section support the results of the
previous section, though the approach is different.

A Hamiltonian analysis of ``beyond Horndeski'' theories was
carried out in
\cite{Gleyzes:2014dya,Gleyzes:2014qga,Lin:2014jga}, though in
that paper the authors restricted their attention to the unitary
gauge, $t=\varphi$. As we shall see below, in the unitary gauge
the Hamiltonian analysis is greatly simplified. Indeed, in an
arbitrary gauge, the Lagrangian (from which the Hamiltonian is
constructed) explicitly contains a term in $\ddot{\varphi}$
multiplied by first time derivatives of the metric: this is the
term which generates third time derivatives in the equations of
motion. However, this term is also multiplied by {\it spatial}
derivatives of $\varphi$ which vanish in the unitary gauge. Hence
in the unitary gauge the Lagrangian contains no second-order time
derivatives. Their presence (in a gauge-invariant calculation, as
discussed here) renders a Hamiltonian analysis much more involved
as we shall see.

In this section we carry out a Hamiltonian analysis for the
non-gauge fixed action
\begin{equation}
S = \int d^4 x \sqrt{-g} \left[R + {\cal L}_{(4,0)}\right],
\label{D:action}
\end{equation}
where we recall that ${\cal L}_{(4,0)}$ was given in
Eq.~(\ref{D:L4}). However, precisely because of the presence of
the $\ddot{\varphi}$ terms mentioned above, the calculation will
quickly become complex and lead to very sizeable expressions.
Thus for reasons of clarity, we will not display all expressions
in their full gory detail: a courageous reader is referred to
Appendix \ref{AppB} for more details. In fact, we shall push the
calculation only as far as to be able to conclude that the theory
necessarily possesses less than 4 Lagrangian degrees of freedom.

\subsection{ADM parametrization and primary constraints}

In the ADM parametrization (with lapse $N$, shift $N^i$, and
spatial metric $\gamma_{ij}$), action (\ref{D:action}) becomes
\begin{eqnarray}
S &=&
\int dt d^3x \, N \sqrt{\gamma} (K_{ij}K^{ij} - K^2 + {}^{(3)}R)
\nonumber
\\
&+& \int dt d^3x \, \frac{\sqrt{\gamma}}{N}
\epsilon^{ijk} \epsilon^{\ell m}{}_k \big[
- \dot{\varphi}^2 s_{i\ell} s_{jm}
-2\varphi_i\varphi_\ell s_{00}s_{jm}
+ 2 \varphi_i \varphi_\ell s_{0m} s_{0j}
+ 4 \dot{\varphi}\varphi_{\ell} s_{i0} s_{jm} \big]
\nonumber
\\
&+& \int dt d^3x \, \frac{\sqrt{\gamma}}{N}
\epsilon^{ijk}\epsilon^{\ell m n} N_k \big[
2 \dot{\varphi} \varphi_{\ell} s_{im}s_{jn}
- 4 \varphi_i \varphi_{\ell} s_{0m} s_{jn} \big]
\nonumber
\\
&+& \int dt d^3x \, N\sqrt{\gamma} \left(1
- \frac{N_p N^p}{N^2}\right)
\epsilon^{ijk}\epsilon^{\ell m n} s_{jm}s_{kn} \varphi_i \varphi_\ell
\label{D:thon}
\\
&\equiv & \int d^4x L,
\label{D:orgal}
\end{eqnarray}
where
\begin{eqnarray}
s_{\mu \nu} &\equiv & \nabla_\mu \nabla_\nu \varphi,
\label{D:Sdef}
\end{eqnarray}
and $K_{ij}$ is the extrinsic curvature (see Appendix
\ref{AppA}).

The first line, linear in the lapse, is the usual ADM
decomposition of the Einstein Hilbert action in General
Relativity (GR). Notice, however, that due to the other terms,
the Lagrangian $L$ is no longer linear either in $N$ nor $N^i$.
Furthermore, ${L}$ generically contains products of second
(covariant) derivatives of $\varphi$ (the $s_{\mu \nu}$), which
in turn contain Christoffel symbols ---~which are expressed in
terms of time derivatives of the lapse and shifts, see Appendix
\ref{AppA}. Thus the action depends explicitly and non-linearly
on $\dot{N}$ and $\dot{N}^i$, as opposed to the the case of GR.
Notice that the term $\propto \varphi_i\varphi_\ell s_{00}s_{jm}$
(mentioned above) in principle generates 3rd order derivatives in
the equations of motion, though it vanishes in the unitary gauge.

More generally, the non-linear dependence of action $S$ on the
variables $s_{\mu \nu}$ makes it very difficult to invert
$\dot{\gamma}_{ij}$ in terms of its conjugate momenta $\pi^{ij}
\equiv {\delta L}/{\delta \dot{\gamma}_{ij}} $. To alleviate this
problem, we proceed by linearizing the action in second
derivatives. [Note that this technique would also work for
higher-order Galileon theories, for instance ${\cal L}_{(5,0)}$.]
That is, we rewrite the action (\ref{D:orgal}) as
\begin{eqnarray}
\tilde{S}
&=& S + \int d^4 x
\; \tilde{\lambda}^{\mu \nu} \left(s_{\mu \nu}
- {\nabla_{\mu} \nabla_{\nu} {\varphi}}\right),
\label{D:orgalbis}
\end{eqnarray}
where $s_{\mu \nu}$ and $\tilde{\lambda}_{\mu \nu}$ are just
symmetric tensors considered as dynamical fields. The field
$\tilde{\lambda}_{\mu \nu}$ is a Lagrange multiplier imposing the
relation (\ref{D:Sdef}). It is straightforward to check that the
equations of motion following from the two actions
(\ref{D:orgal}) and (\ref{D:orgalbis}) are equivalent. Notice,
however, that the price to pay for this linearization is the
introduction of new degrees of freedom: indeed the dynamical
fields are now
\begin{equation}
N\, ,N^i \,,\gamma_{ij}\, , \varphi, \lambda_{\mu \nu}\, , s_{\mu \nu},
\label{D:var}
\end{equation}
where, for computational simplicity we choose to work with
\begin{equation}
\lambda^{\mu \nu} = N\sqrt{-\gamma} \tilde{\lambda}^{\mu \nu}
\end{equation}
rather than $\tilde{\lambda}^{\mu \nu}$. Thus there are a total
of 31 dynamical fields, together with their 31 conjugate momenta
defined by
\begin{eqnarray}
&& \pi_N \equiv \frac{\delta L}{\delta \dot{N}} \,, \qquad
\pi_{i} \equiv \frac{\delta L}{\delta \dot{N}^i}
\, , \qquad \pi^{ij} \equiv \frac{\delta L}{\delta \dot{\gamma}_{ij}},
\nonumber
\\
&& \pi_\varphi \equiv \frac{\delta L}{\delta \dot{\varphi}} \, , \qquad
\pi^{(\lambda)}_{\mu \nu} \equiv
\frac{\delta L}{\delta \dot{\lambda}^{\mu \nu}} \, , \qquad
\pi_{(s)}^{\mu \nu} \equiv\frac{\delta L}{\delta \dot{s}_{\mu \nu}} .
\end{eqnarray}
The fields and their conjugate momenta satisfy the standard
Poisson-Bracket (PB) relations, for instance $\{N(x), \pi_N(y) \}
= \delta^3(x,y)$ (see Appendix \ref{AppB} for the remaining
---~obvious~--- relations).

The canonical momenta are determined directly from (\ref{D:thon})
and (\ref{D:orgalbis}), and we find
\begin{equation}
\pi_{(s)}^{\mu \nu} =0 \,, \qquad \pi^{(\lambda)}_{0 i}
= 0 \,, \qquad \pi^{(\lambda)}_{ij} = 0 \,,
\qquad \pi_i = \lambda^{00} \varphi_i \, ,\qquad \pi_N =
\frac{1}{N}\left(\lambda^{00} \pi^{\lambda}- N^i \pi_i\right),
\label{D:con}
\end{equation}
where
\begin{equation}
\pi^{\lambda} \equiv \pi^{(\lambda)}_{0 0} = \dot{\varphi} .
\label{D:pilambda}
\end{equation}
The remaining conjugate momenta $\pi_\varphi$ and $\pi^{ij}$ are
given in Appendix \ref{AppB}, equations (\ref{D:pphi}) and
(\ref{D:ppg}) respectively. In particular $\pi_\varphi$ depends
linearly on both $\dot{\lambda}^{00}$ and $\dot{N}$ [actually on
$\partial_0(\lambda^{00} N)$]. Hence, given the expression for
$\pi^\lambda$ in (\ref{D:pilambda}), naively one might expect
this theory to contain two scalar degrees of freedom,
$\lambda^{00}$ and $\varphi$. The momentum $\pi^{ij}$ conjugate
to $\gamma_{ij}$ depends on $\dot{\gamma}_{ij}$ (as expected) as
well as $\dot{\varphi}$.

The 23 relations in (\ref{D:con}) define 23 primary constraints
\begin{eqnarray}
\Phi_{(s)}^{\mu \nu} &\equiv& \pi_{(S)}^{\mu \nu} \approx 0,
\label{D:p1}
\\
\Phi^{(\lambda)}_{0i} &\equiv& \pi^{(\lambda)}_{0i}\approx 0,
\\
\Phi^{(\lambda)}_{ij} &\equiv& \pi^{(\lambda)}_{ij}\approx 0,
\\
\Phi_i &\equiv& \pi_i - \lambda^{00} \varphi_i \approx 0,
\label{D:phii}
\\
\Phi_N &\equiv&\pi_N - \frac{\lambda^{00}}{N}
\left(\pi^{\lambda}- N^i \varphi_i\right) \approx 0.
\label{D:listP}
\end{eqnarray}
It is straightforward to see that on shell, {\it all} the primary
constraints have vanishing Poisson brackets amongst each other.
For the following discussion, it will be useful to denote the
primary constraints by
\begin{equation}
\{\Phi_{(s)}^{00}, \Phi_i ,\Phi_N,\Phi_{\tilde P} \},
\qquad {\tilde P}=1,\ldots,18.
\label{D:primary}
\end{equation}
That is we separate out $\Phi_{(s)}^{00}$, $\Phi_i$ and $\Phi_N$
{}from the remaining $\Phi_{\tilde P}$ primary constraints,
since, as we shall see below, they must be considered differently
{}from the others. Recall that in GR, $\Phi_i$ and $\Phi_N$
correspond, respectively, to the primary constraints $\pi_i
\approx 0$, $\pi_N \approx 0$ which constitute 4 of the 8 first
class constraints associated with diffeomorphism invariance. In a
similar way, 4 of the primary constraints in (\ref{D:primary})
(namely, a linear combination of $\Phi_i$ and $\Phi_N$ with the
other primary constraints) must also be first-class constraints.
There must also be 4 secondary constraints of first class
---~which, in turn, do not generate tertiary or higher generation
constraints~--- as a consequence of the fact that diffeomorphism
invariance is expressed infinitesimally through four independent
parameters (the four components of a vector) which are just
differentiated once with respect to time
\cite{Henneaux:1990au,Henneaux:1989zc}.

\subsection{Canonical Hamiltonian and secondary constraints}

The remaining Hamiltonian analysis follows the standard route
(see for instance \cite{Henneaux:1992ig}), but is rather involved
due to the large number of fields and the intrinsically
non-linear nature of the problem. The first step is to calculate
the canonical Hamiltonian $H_c$ from which, as a second step we
determine the secondary constraints, obtained by imposing the
preservation of primary constraints under time evolution. We find
\begin{eqnarray}
H_c&=& \int d^3 x \Big \{N\sqrt{\gamma}\left[
{(K_{ij}K^{ij} - K^2)} - {}^{(3)}R \right]
+ 2 (D_i N_j)\pi^{ij} - {\lambda^{\mu \nu}}s_{\mu \nu}
+ (\pi_\varphi\pi^{\lambda})
\nonumber
\\
&& \qquad \qquad - (P \pi^{\lambda} + q) \pi^{\lambda}- V
\nonumber
\\&& \qquad \qquad+ (D_i D_j \varphi) ({\lambda^{ij}}
+ 2{\lambda^{0i}} N^j) -{\lambda^{00}}
\varphi^k \left[ N^q (D_q N_k) + N (\partial_k N) \right]
\nonumber
\\
&& \qquad \qquad -
\pi_N \left[ N^k (\partial_k N) + \frac{2}{\lambda^{00}}
\partial_i (N {\lambda^{0i}}) \right] \Big\},
\label{D:Hcanonical}
\end{eqnarray}
where $P$, $q$ and $V$ are various combinations of functions
appearing in the action (\ref{D:thon}), as defined below. To
simplify expressions, we define
\begin{equation}
{\cal F}^{i\ell jm} = \epsilon^{ijk} \epsilon^{\ell m}{}_k
\label{D:Fdef}
\end{equation}
so that
\begin{eqnarray}
P &=& - \frac{\sqrt{\gamma}}{N} {\cal F}^{i\ell jm}s_{i\ell} s_{jm},
\label{D:pipi}
\\
q &=& \frac{2\sqrt{\gamma}}{N} \varphi_{\ell}
\left( 2 {\cal F}^{i\ell jm} s_{i0} s_{jm}
+ \epsilon^{ijk}\epsilon^{\ell m n}N_k s_{im}s_{jn}\right),
\label{D:ququ}
\\
V&=& \frac{\sqrt{\gamma}}{N} 2 \varphi_\ell \varphi_i
\left[{\cal F}^{i\ell jm}\left( s_{0m} s_{0j}- s_{00}s_{jm} \right)
- 2 \epsilon^{ijk}\epsilon^{\ell m n} N_k s_{0m} s_{jn} \right]
\nonumber\\
&& + N \sqrt{\gamma} \left(1 - \frac{N_p N^p}{N^2}\right)
\epsilon^{ijk}\epsilon^{\ell m n} s_{jm}s_{kn} \varphi_i \varphi_\ell,
\label{D:Vdef}
\end{eqnarray}
and the extrinsic curvature $K_{ij}$ is given in terms of the
canonical momenta $\pi^{ij}$ by
\begin{equation}
\sqrt{\gamma} K^{ij} = \Lambda^{ij} - \gamma^{ij}\frac{\Lambda}{2},
\end{equation}
where
\begin{eqnarray}
\Lambda^{ij} &=& \pi^{ij} - \frac{\pi_N}{2N} \left( N^i N^j
+ {\frac{\lambda^{ij}}{\lambda^{00}}}
+ 2{\frac{\lambda^{0(i} N^{j)}}{\lambda^{00}}} \right)
- \left({\lambda^{0(j}} \varphi^{i)}
+ {\lambda^{00}} N^{(j}\varphi^{i)}\right).
\label{Lambdadef}
\end{eqnarray}
Notice that, because of the term linear in $\pi_\varphi$, the
canonical Hamiltonian appears at first sight not to be bounded
{}from below, \`a la Ostrogradski \cite{Ostrogradski}. However,
we shall see that $\pi_\varphi$ is in fact not independent of the
other fields, because of one of the secondary constraints (to be
precise it is the constraint ${\cal H}_0$, defined in
(\ref{D:H0}) and given explicitly in equation (\ref{H0}) in
Appendix \ref{AppB}). As a result the canonical Hamiltonian will
actually vanish on shell, as expected for a
diffeomorphism-invariant theory, see Eq.~(\ref{D:Hc}). Following
the standard procedure, the total Hamiltonian is then given by
\begin{equation}
H_T = H_c + \int d^3 x \left[
\zeta^{(s)}_{\mu \nu}\Phi_{(s)}^{\mu \nu}
+ \zeta_{(\lambda)}^{0i} \Phi^{(\lambda)}_{0i}
+ \zeta_{(\lambda)}^{ij}\Phi^{(\lambda)}_{ij}
+ \zeta^i \Phi_i + \zeta_N \Phi_N \right],
\label{D:HT}
\end{equation}
thus introducing 23 Lagrange multipliers, $\zeta$.

Imposing the conservation of the primary constraints,
schematically $ \dot{\Phi}(x) = \{\Phi(x),H_T \}$, enables us to
determine 23 corresponding secondary constraints (denoted by
$\kappa = \dot{\Phi}(x)$). Using Eq.~(\ref{D:Hcanonical})
together with the primary constraints in (\ref{D:listP}), we find
\begin{eqnarray}
\kappa^{00}_{(S)} &=& -2\frac{\sqrt{\gamma}}{N}
{\cal F}^{i\ell jm} \varphi_i \varphi_\ell s_{jm} + \lambda^{00},
\label{D:kappa00}
\\
\kappa^{0i}_{(S)} &=& 4 \frac{\sqrt{\gamma}}{N}
\left( \pi^{\lambda} {\cal F}^{i \ell jk}s_{jk}
\varphi_\ell + {\cal F}^{k\ell j i} s_{0j} \varphi_k \varphi_\ell
- \epsilon^{qjk}\epsilon^{\ell i n} N_{k} s_{jn}
\varphi_{\ell} \varphi_{q}\right) + 2 \lambda^{0i} ,
\label{D:thisone}
\\
\kappa^{pq}_{(S)} &=& \lambda^{pq} -2\frac{\sqrt{\gamma}}{N}
(\pi^{\lambda})^2 {\cal F}^{pqjm}s_{jm}
+4 \frac{\sqrt{\gamma}}{N} \pi^\lambda
\left\{{\cal F}^{i\ell (pq)} \varphi_\ell s_{i0}
- \epsilon^{jk(p}\epsilon^{q)\ell n} N_k \varphi_\ell s_{jn}\right\}
\nonumber
\\
&& - 2\frac{\sqrt{\gamma}}{N} {\cal F}^{i\ell pq}
\varphi_i \varphi_\ell s_{00}
+ 2 \frac{\sqrt{\gamma}}{N} \epsilon^{ij(p}\epsilon^{q)\ell m}
\varphi_i \varphi_\ell \left\{2 N_j s_{0m}
+ \left(N^2-{N_fN^f}\right)s_{jm} \right\},
\label{D:kappaSpq}
\\
\kappa^{(\lambda)}_{ij} &=& s_{ij} - D_i D_j \varphi
+ \frac{K_{ij} \pi_N}{\lambda^{00}},
\label{D:gosh}
\\
\nonumber
\\
\kappa^{(\lambda)}_{0i} &=& 2s_{0i} - 2 (D_i D_j \varphi)N^j
- 2 N \partial_i \left(\frac{\pi_N}{\lambda^{00}}\right)
+ 2 K_{ij} \left[ N^j\frac{\pi_N}{\lambda^{00}} + N\varphi^j \right] .
\label{D:kappaLoi}
\end{eqnarray}
The last 4 secondary constraints, denoted by
\begin{equation}
{\cal H}_0 = - \{\Phi_N, H_T\} \, ,
\qquad {\cal H}_i = - \{\Phi_i, H_T\},
\label{D:H0}
\end{equation}
in analogy with the Hamiltonian and momentum constraints in GR,
are given in Appendix \ref{AppB}. The first, ${\cal H}_0$
contains a term in $\pi_\varphi \pi_N$, while the second $ {\cal
H}_i $ contains a term in $D_j \pi^j_{\; \; i}$. In linear
combinations with the other (primary and secondary) constraints,
they must therefore constitute the 4 remaining secondary
first-class generators associated with diffeomorphism invariance.

In analogy with the primary constraints, it will be useful to
write the set of secondary constraints as
\begin{equation}
\{\kappa^{(s)}_{00}, {\cal H}_i ,{\cal H}_0 ,
\kappa_{\tilde P} \}, \qquad {\tilde P}=1,\ldots,18.
\label{D:secondary}
\end{equation}
Finally, in terms of the primary and secondary constraints, the
canonical Hamiltonian in Eq.~(\ref{D:Hcanonical}) can be
expressed as
\begin{eqnarray}
H_c&=& \int d^3 x \Big\{N {\cal H}_0
+ N_i {\cal H}^i - \left(2\kappa^{00}_{(s)} s_{00}
+ \kappa^{0i}_{(s)}s_{0i}\right) - \kappa_{ij}^{(\lambda)} \lambda^{ij}
\nonumber
\\
&+& {\Phi_N}\left[ K_{ij} \left( N^i N^j
+ 2\frac{\lambda^{0(i} N^{j)}}{\lambda^{00}}
+ \frac{\lambda^{ij}}{\lambda^{00}} \right)
+ \frac{2}{{\lambda_{00}}} \partial_i (N\lambda^{0i})
+ N^i \partial_i N \right] \Big\}.
\label{D:Hc}
\end{eqnarray}
That is, as expected for a diffeomorphism-invariant theory, the
Hamiltonian vanishes on shell.

\subsection{Counting degrees of freedom, first and second-class
constraints}
\label{Sec3C}

The theory has $2 \times 31 = 62$ Hamiltonian degrees of freedom,
and so far we have identified 23 primary and 23 secondary
constraints. Of these, at least 8 must be first class (due to
diffeomorphism invariance). {\it If} all the remaining
constraints were second class and {\it if} there were no tertiary
constraints, then at this stage we would conclude that the theory
contains
\begin{equation}
62 - (2\times 8) - (46-8) = 8
\label{D:count}
\end{equation}
Hamiltonian degrees of freedom. That is 4 Lagrangian degrees of
freedom: 2 for the graviton and 2 scalars.

The next step therefore consists in determining whether or not
the remaining primary and secondary constraints are of second
class. To do so, as per the standard procedure, one must
calculate the $46 \times 46$ matrix of their Poisson brackets. We
order the constraints with the primary constraints first followed
by the secondary ones. At this stage it is useful to notice that
this $46 \times 46$ matrix is of the anti-diagonal form
\begin{equation}
\left(
\begin{array}{rcc}
\mathbf{0} & &{\cal A}\\
-{\cal A} & &{\cal B}
\end{array}
\right),
\end{equation}
where first $23\times23$ block vanishes since the primary
constraints all commute (see above), and the $23\times23$ block
${\cal A}$ is itself symmetric because of the Jacobi identity:
\begin{equation}
\{\Phi_i,\kappa_j \} = \{\Phi_i, \{\Phi_j,H_T \} \} =
- \{\Phi_j, \{H_T,\Phi_i \} \}
- \{H_T, \{\Phi_i,\Phi_j \} \} =+ \{\Phi_j,\kappa_i \}.
\end{equation}

We now focus on a subset of the primary and secondary
constraints, namely the $\Phi_{\tilde{P}}$ and
$\kappa_{\tilde{P}}$; the remaining constraints will be discussed
later. The Poisson brackets of these constraints define a
$36\times36$ matrix of the same anti-diagonal form, where the
antidiagonal block is the 18$\times 18$ matrix
\begin{equation}
{\cal M}_{18} \equiv \{\Phi_{\tilde{P}},
\kappa_{\tilde{Q}} \}, \qquad (\tilde{P},\tilde{Q})=1,\ldots,18.
\label{M18}
\end{equation}
The determinant of this $36\times36$ matrix is $=-(\det({\cal
M}_{18}))^2$, and using the commutators given in appendix
\ref{AppB} one can check that det$({\cal M}_{18})\neq 0$. Hence
these $18\times2=36$ constraints are necessarily second class.
This also implies that they do not generate any tertiary
constraints, because $\{\kappa,H_T \} \approx 0$
actually fix the
corresponding Lagrange multipliers $\zeta$ entering the total
Hamiltonian in Eq.~(\ref{D:HT}).

However, as soon as we consider the 19$\times 19$ matrix
consisting of $M_{18}$ but augmented by the further primary
constraint $\Phi^{00}_{(s)}$, and the further secondary
constraint $\kappa^{(s)}_{00}$, the determinant vanishes. Hence
there exists a linear combination of $\Phi^{00}_{(s)}$ and
$\Phi_{\tilde{P}}$ which commutes with all the
$(\Phi^{00}_{(s)},\Phi_{\tilde{P}},\kappa^{(s)}_{00},
\kappa_{\tilde{P}})$. The explicit expression for this linear
combination, denoted by $\tilde{\Phi}^{00}_{(s)}$, is given in
Appendix \ref{AppB}, equation (\ref{D:tildePhi}). Similarly we
have shown that there exists a linear combination, denoted by
$\tilde{\kappa}^{(s)}_{00}$, of $\kappa^{(s)}_{00}$ with the
$\kappa_{\tilde{P}}$'s and the primary constraints which commutes
with all the
$(\Phi^{00}_{(s)},\Phi_{\tilde{P}},\kappa^{(s)}_{00},
\kappa_{\tilde{P}})$.

At this stage, the Hamiltonian analysis of ${\cal L}_4$ must
necessarily proceed in one of the two following ways:
\begin{enumerate}
\item {\it Either} $\tilde{\Phi}^{00}_{(s)}$ and
$\tilde{\kappa}^{(s)}_{00}$ also commute with the remaining
constraints (namely ${\cal H}_0$, ${\cal H}_i$, $\Phi_N$ and
$\Phi_i$), and hence constitute two more first-class constraints
in addition to the $2\times4$ implied by diffeomorphism
invariance. In this case the theory has 6 Hamiltonian degrees of
freedom (2 for the graviton and 1 for the scalar field).

\item {\it Or} $\tilde{\Phi}^{00}_{(s)}$ and
$\tilde{\kappa}^{(s)}_{00}$ do {\it not} commute with the
remaining constraints (namely ${\cal H}_0$, ${\cal H}_i$,
$\Phi_N$ and $\Phi_i$). In that case there exist a tertiary and
possibly quaternary (or higher) constraints. Indeed, the
vanishing of the above $19\times 19$ determinant implies that the
Lagrange multiplier $\zeta_{00}^{(s)}$ entering (\ref{D:HT}) is
not fixed by the conservation of the secondary constraint,
$\{\tilde{\kappa}^{(s)}_{00}, H_T\} \approx 0$.
\end{enumerate}
Whichever is correct, we arrive at the important conclusion that
there are necessarily \textit{less} than 8 Hamiltonian degrees of
freedom.

In order to see whether the first option is the true one, we note
that if they are indeed first-class generators,
$\tilde{\Phi}^{00}_{(s)}$ and $\tilde{\kappa}^{(s)}_{00}$ must
generate a new symmetry. However, we have determined the
transformations of the fields induced by these generators, and
found that the action of the theory in \textit{not} invariant.
Hence, necessarily we arrive at the conclusion that case 2 is
correct: there must exist a tertiary (and even perhaps a
quaternary) constraint beyond the primary and secondary
constraints found above. These are generated by the conservation
of $\tilde{\kappa}^{(s)}_{00}$. Unfortunately, though, because of
the size of the corresponding Poisson brackets, the computation
of these tertiary/quaternary constraints is a cumbersome task we
felt unnecessary to attempt.

However, we have carried out an identical calculation in the
simpler case of ${\cal L}_{(2,0)}$ (using the \textit{same}
technique with our Lagrange multipliers~(\ref{D:orgalbis}),
although of course this greatly complicates the calculation in
this very simple case). Here we have shown that there are indeed
a tertiary constraint generated by the conservation of
$\tilde{\kappa}_{00}^{(s)}$, and even a quaternary constraint
generated by the conservation of this tertiary one. There are no
higher order constraints because the conservation of the
quaternary constraints actually imposes the value of the Lagrange
multiplier $\zeta_{00}^{(s)}$. It happens that the secondary and
tertiary do not commute with each other, and that the primary and
the quaternary also do not commute with each other. Therefore,
all of these four constraints are actually of second class. As
compared to Eq.~(\ref{D:count}), there are thus 2 extra
constraints, which give a final number of 6 Hamiltonian degrees
of freedom. That is, 3 Lagrangian degrees of freedom: 2 for the
graviton and a single scalar. We expect the same result to hold
true also for ${\cal L}_{(4,0)}$, although we did not do this
explicit calculation because of its complexity, and due to the
conclusions of Sec.~\ref{Sec2} which we felt rendered it
unnecessary.

\section{Conclusions}
In this work, we have considered Galileon models in $D$
dimensions, as well as the models defined by the counterterms
which have been introduced to maintain the second-order nature of
the field equations when both the metric and the scalar are made
dynamical. We have first shown that in one given such model, all
the third time derivatives which appear in the field equations
can be eliminated, leaving a set of $\frac{D(D+1)}{2}+1$ field
equations with at most second time derivatives of the dynamical
fields. The same has been shown to hold for an arbitrary linear
combination of such models, as well as their k-essence-like
generalizations involving free functions of $\varphi$ and
$\varphi_\lambda^2$. (In $D$ dimensions, these models can depend
on $\lfloor\frac{D+1}{2}\rfloor\lfloor\frac{D}{2}+1\rfloor$
independent such functions.) This supports the claim made
previously \cite{Gleyzes:2014dya,Gleyzes:2014qga} that the number
of degrees of freedom in these theories is only 3, counting 2 for
the graviton and 1 for the scalar. However, it does not provide
an absolutely rigorous proof of this claim which would require a
detailed Hamiltonian analysis of these models. Such a Hamiltonian
analysis has been carried out in the second part of this paper
for one of the models under consideration, reaching the conclusion
that the number of degrees of freedom is indeed strictly less
than 4. It seems, however, very likely, in light of our results,
and taking into account the gauge invariance of the theory (which
has been fully kept in our analysis, in contrast to previous
works) that the final number of degrees of freedom is well only
3. Indeed, using the first part of this paper, one expects that
the ``reduced'' field equations (containing only second
derivatives) can still be decomposed into 4 Lagrangian
constraints (coming from the invariance of the theory under
reparametrization) plus 6 dynamical equations for 6 dynamical
metric variable and an other one for the scalar. Note, however,
that the extraction of these would be 4 Lagrangian constraints is
much trickier than in standard general relativity because it can
be checked that, in an ADM language, the field equations contain
second time derivatives of the spatial part of the metric, but
also of the lapse and the shift. Provided these Lagrangian
constraints exist, the gauge invariance would then reduce to two
the number of dynamical components in the metric. Rigorously
speaking, our Hamiltonian analysis also leaves open the
possibility that there is just one tertiary second class
constraint (and no quaternary). This would result in an odd
(times infinity\footnote{In classical mechanics, there is always
an even number of second-class constraints, but this is no longer
true in continuous field theories, where each constraint should
actually be understood as an infinity of them, since it is
imposed at every point of the Cauchy surface.}) number of second
class constraints, which can happen in a field theory (see e.g.
\cite{Henneaux:2009zb} and references therein), but seems to us
unlikely in a bosonic and Lorentz-invariant theory. This,
however, deserves further investigation to be rigorously checked.

\section*{Acknowledgments}
The work of C.D.~was supported by the European Research Council
under the European Community's Seventh Framework Programme
(FP7/2007-2013 Grant Agreement no.~307934, NIRG project). The
work of G.E.F.~was partially supported by the ANR THALES grant.
D.A.S.~is grateful to CERN for hospitality whilst this work was
in progress, and acknowledges the support of the excellence
cluster/Labex ENIGMASS. D.A.S.~thanks K.~Noui for useful
discussions on the Hamiltonian analysis, and dedicates this work
to little Sebastian M\"ostl who has been an inspiration
throughout.

\appendix
\section{Christoffel symbols in ADM variables}
\label{AppA}

In the ADM notation, the 4D metric and its inverse are written as
\begin{eqnarray}
g_{\mu \nu} = \left(\begin{array}{cc} -N^2 + N_kN^k & N_i \\
N_j & \gamma_{ij}
\end{array} \right) \, , \qquad &&
g^{\mu \nu} = \left(\begin{array}{cc}
-\frac{1}{N^2} & \frac{N^i}{N^2} \\
\frac{N^j}{N^2} & \gamma^{ij}- \frac{N^i N^j}{N^2}
\end{array} \right),
\label{ADM}
\end{eqnarray}
where here and in the following all spatial indices $i,j,k, ...$
are raised and lowered with the spatial metric $\gamma_{ij}$.
On using (\ref{ADM}) the Christoffel symbols can be calculated.
These will contain $\dot{\gamma}_{ij}$, $\dot{N}$ and
$\dot{N}^i$. The covariant derivative associated with the metric
$\gamma_{ij}$ is denoted by $D$, and $K_{ij}$ is defined by
\begin{eqnarray}
K_{ij} = \frac{1}{2N}\left(\dot{\gamma}_{ij}
- D_i N_j- D_j N_i \right).
\end{eqnarray}
We find
\begin{eqnarray}
\Gamma^0_{00} &=&
\frac{1}{N} \left(\dot{N}+ N^i \partial_i N +K_{ij}N^i N^j \right),
\nonumber \\
&&
\nonumber
\\
\Gamma^j_{00} &=& -\frac{\dot{N}}{N}N^j
+ \dot{N}^j + 2 N N^q K_{qk} \left( \gamma^{jk}
- \frac{N^k N^j}{2N^2} \right) + N^q \gamma^{jk}(D_q N_k)
\nonumber
\\
&&
+ N \partial_k N
\left( \gamma^{jk} - \frac{N^j N^k}{N^2}\right),
\nonumber \\
\Gamma^0_{j0}
&=& \frac{1}{N}\left( \partial_j N+ K_{jl} N^l \right),
\nonumber
\\
\Gamma^l_{j0}
&=& -\frac{N^l \partial_j N}{N} - \frac{1}{N} N^l N^m K_{jm}
+ \gamma^{lm} K_{jm} N + \gamma^{lm} D_j N_m,
\nonumber \\
&&
\nonumber
\\
\Gamma^0_{jl}
&=& \frac{1}{N} K_{jl},
\nonumber \\
&&
\nonumber
\\
\Gamma^n_{jl}
&=& -\frac{N^n}{N} K_{jl} + \Gamma^n_{jl}(\gamma),
\end{eqnarray}
where $\Gamma^n_{jl}(\gamma)$ stands for the Christoffel symbols
of the spatial metric $\gamma_{ij}$.

\section{Hamiltonian analysis of ${\cal L}_{(4,0)}$}
\label{AppB}

\noindent
We give here some of the intermediate expressions required for
Sec.~\ref{Sec3}. We will use one abuse of notation in this
Appendix (and only here): namely we denote $\gamma^{ij}
\varphi_j$ by $\varphi^i$.

The 31 dynamical variables given in (\ref{D:var}), namely
\begin{equation}
N\, ,N^i \,,\gamma_{ij} \, \, ,\, \,
\varphi, \lambda_{\mu \nu}\, , s_{\mu \nu},
\end{equation}
and their conjugate momenta
\begin{equation}
\pi_N \, ,\pi_{i}\,, \pi^{ij} \, \, , \, \,
\pi_\varphi, \pi^{(\lambda)}_{\mu \nu}\, , \pi_{(s)}^{\mu \nu},
\end{equation}
satisfy the Poisson Brackets
\begin{eqnarray}
\{N(x), \pi_N(y) \} &=& \delta^3(x,y) \, ,
\nonumber
\\
\{N^i(x), \pi_j(y) \} &=& \delta^{i}_j \delta^3(x,y) \, ,
\nonumber
\\
\{\gamma_{ij}(x), \pi^{pq}(y) \} &=&
\delta_{(i}^p \delta_{j)}^q \delta^3(x,y) \, ,
\nonumber
\\
\{\lambda^{00}(x), \pi^\lambda_{00}(y) \} &=&
\delta^3(x,y) \, ,\qquad \qquad\{s_{00}(x),
\pi_S^{00}(y) \} = \delta^3(x,y) \, ,
\nonumber
\\
\{\lambda^{0i}(x), \pi^\lambda_{0j}(y) \} &=&
\delta^{i}_j \delta^3(x,y) \, , \, \qquad \; \; \; \;
\{s_{0i}(x), \pi_S^{0j}(y) \} = \delta^{j}_i \delta^3(x,y) \, ,
\nonumber
\\
\{\lambda^{ij}(x), \pi^\lambda_{pq}(y) \} &=&
\delta^{(i}_p \delta^{j)}_q \delta^3(x,y) \, \, ,\qquad
\{s_{ij}(x), \pi_S^{pq}(y) \} =
\delta_{(i}^p \delta_{j)}^q \delta^3(x,y) \, \, ,
\end{eqnarray}
with all other commutators vanishing.

Starting from (\ref{D:thon}), (\ref{D:orgal}) and
(\ref{D:orgalbis}), the conjugate momenta $\pi_\varphi$ and
$\pi^{ij} $ are found to be given by
\begin{eqnarray}
\pi_\varphi &=& \dot{\lambda}^{00} + \hat{Q}
+ 2 P \pi^{\lambda} + \frac{\lambda^{00}\dot{N}}{N}
+ \frac{K_{ij} Q^{ij}}{N},
\label{D:pphi}
\\
\pi^{ij} &=& \sqrt{\gamma}(K^{ij} - K\gamma^{ij})
+ \frac{\varphi_m T^{mij}}{2N} + \frac{Q^{ij}\dot\varphi}{2N^2},
\label{D:ppg}
\end{eqnarray}
where
\begin{eqnarray}
{\hat{Q}} &=& \frac{1}{N} (\partial_i N) N^i {\lambda^{00}}
+ \frac{2}{N} \partial_i \left( N {\lambda^{0i}}\right),
\nonumber
\\
{Q^{ij}} &=&{\lambda^{00}}N^i N^j + 2 \lambda^{0i} N^j + \lambda^{ij},
\nonumber
\\
{T^{imn}} &=& 2NN^n {\lambda^{00}} \left(\gamma^{im}
- \frac{N^i N^m}{2N^2} \right)
+ 2 \lambda^{0n} N \left(\gamma^{im} - \frac{N^i N^m}{N^2} \right)
-\lambda^{mn}\frac{N^i}{N}.
\end{eqnarray}

On using (\ref{D:phii}), (\ref{D:listP}) and
(\ref{D:Hcanonical}), one can calculate the secondary constraints
\begin{eqnarray}
{\cal H}_0 &\equiv& - \{\Phi_N, H_T\},
\nonumber
\\
{\cal H}_i &\equiv& - \{\Phi_i, H_T\}.
\end{eqnarray}
The first is given by
\begin{eqnarray}
{\cal H}_0
&=&
\sqrt{\gamma} \left[ (K_{ij}K^{ij} - K^2) - {}^{(3)}R \right]
+ K_{ij}B^{ij}
+ \frac{\pi_\varphi \pi_N}{\lambda^{00}} + \frac{\lambda^{00}s_{00}}{N}
\nonumber
\\
&&- \frac{1}{N}\left[ {P} (\pi^\lambda)^2
+ {(- V_1 + V_2)} - \left({N^i \varphi_i}\right)
(2\pi^\lambda P + q) \right]
\nonumber
\\
&&+ 2 \lambda^{0i} \partial_i \left(\frac{\pi_N}{\lambda^{00}}\right)
+ \partial_i\left(N \lambda^{00} \varphi^i + \pi_N N^i \right)
+ \frac{\pi_N}{N} N^i (\partial_i N)
\nonumber
\\
&& + \frac{\lambda^{00}}{N} \varphi^i N^j (D_j N_i)
- (\partial_i \pi^\lambda) \left(\frac{\lambda^{00}N^i}{N}\right)
\nonumber
\\
&&
-\frac{{\Phi_N}}{N}\left[ K_{ij}\left( N^i N^j
+ 2\frac{\lambda^{0(i} N^{j)}}{\lambda^{00}}
+ \frac{\lambda^{ij}}{\lambda^{00}} \right)
+ N^k \partial_k N
+ 2 \frac{1}{\lambda^{00}}\partial_i (N\lambda^{0i})
+ \frac{N \pi_\varphi}{\lambda^{00}} \right],
\label{H0}
\end{eqnarray}
where
\begin{equation}
B_{ij} = \frac{\pi_N}{N} \left( 2 N^i N^j
+ 2\frac{\lambda^{0(i} N^{j)}}{\lambda^{00}}
+ \frac{\lambda^{ij}}{\lambda^{00}} \right)
+ 2 \lambda^{00} N^{(j}\varphi^{i)},
\label{Bdef}
\end{equation}
and $P$ and $q$ were defined in (\ref{D:pipi}) and (\ref{D:ququ})
respectively. The quantities $V_1$ and $V_2$ are different
components (depending on their $N$-dependence) of $V$ defined in
Eq.~(\ref{D:Vdef})). Namely $V=V_1+V_2$ with
\begin{eqnarray}
V_1 &=& \frac{2 \sqrt{\gamma}}{N} \phi_\ell \phi_i
\left[{\cal F}^{i\ell jm}\left( s_{0m} s_{0j}- s_{00}s_{jm} \right)
- 2 \epsilon^{ijk}\epsilon^{\ell m n} N_k s_{0m} s_{jn} \right]
\nonumber
\\
&&
\qquad
- \frac{\sqrt{\gamma}}{N} {N_p N^p}
\epsilon^{ijk}\epsilon^{\ell m n} s_{jm}s_{kn} \phi_i \phi_\ell,
\label{V1}
\\
V_2 &=& N {\sqrt{\gamma}}
\epsilon^{ijk}\epsilon^{\ell m n} s_{jm}s_{kn} \phi_i \phi_\ell.
\label{V2}
\end{eqnarray}
Finally,
\begin{eqnarray}
{\cal H}_i(x)
&=& -2 D_j \pi^j_{\; \; i} - 2N K_{ij}
\left[ \frac{\pi_N}{N}\left(N^j
+\frac{\lambda^{0j}}{\lambda^{00}}\right)
+ \lambda^{00}\varphi^j \right]
\nonumber
\\
&+& 2 D_j \left( \lambda^{00} N^j \varphi_i \right)
+ 2 (D_j D_i \varphi)\lambda^{0j} - \lambda^{00} \varphi^j(D_i N_j)
- \pi_N \partial_i N + \lambda^{00}\partial_i \pi^\lambda
\nonumber
\\
&+& \varphi_i \pi_\varphi - \varphi_i(2P\pi^\lambda+q)
\nonumber
\\
&+&\frac{2\sqrt{\gamma}}{N} \epsilon^{\ell m n}\varphi_{\ell}
\left[ N_i \left( \epsilon^{q j k} s_{jm} s_{kn} \varphi_q \right)
- \epsilon^{q j}_{\; \; \; \; i} s_{jn}
\left( \pi^\lambda s_{qm} - 2\varphi_{q} s_{0m} \right) \right].
\end{eqnarray}

We now list the different commutators of primary and secondary
constraints, which can be straightforwardly determined from the
primary constraints given in (\ref{D:p1})-(\ref{D:listP}) and the
secondary constraints given in
(\ref{D:kappa00})-(\ref{D:kappaLoi}). In particular, we find that
$ \Phi^{00}_{(s)}$ satisfies
\begin{eqnarray}
\{\Phi^{00}_{(s)}, \kappa^{ij}_{(s)} \} &=&
\frac{2\sqrt{\gamma}}{N} {\cal F}^{ij p q}
\varphi_p \varphi_q = \{\Phi^{ij}_{(s)} ,
\kappa^{00}_{(s)} \},
\end{eqnarray}
with all its remaining commutators (including with the
$\kappa^{(\lambda)}$) vanishing. Then we find
\begin{eqnarray}
\{\Phi^{0i}_{(s)}, \kappa^{0q}_{(s)} \} &=&
-\frac{4\sqrt{\gamma}}{N} {\cal F}^{k\ell q i}
\varphi_k \varphi_\ell,
\nonumber\\
\{\Phi^{0i}_{(s)}, \kappa^{pq}_{(s)} \} &=&
-\frac{4\sqrt{\gamma}}{N}\varphi_\ell \left[
{\cal F}^{i\ell(pq)} \pi^\lambda
+ \epsilon^{fj(p} \epsilon^{q)\ell i}\varphi_f N_j \right],
\nonumber
\\
\{\Phi^{0i}_{(s)}, \kappa_{0j}^{(\lambda)} \} &=& -2 \delta^{i}_j,
\nonumber
\\
\{\Phi^{0i}_{(s)}, \kappa_{pq}^{(\lambda)} \} &=& 0,
\nonumber
\\
\{\Phi^{ij}_{(s)}, \kappa_{pq}^{(\lambda)} \} &=&
- \delta^{i}_{(p} \delta^{j}_{q)},
\nonumber
\\
\{\Phi^{ij}_{(s)}, \kappa^{pq}_{(s)} \} &=&
\frac{2\sqrt{\gamma}}{N} \left[ (\pi^\lambda)^2 {\cal F}^{(pq)(ij)}
+ 2 \pi^\lambda \epsilon^{ik(p} \epsilon^{q)\ell j}\varphi_\ell N_k
- \epsilon^{fj(p} \epsilon^{q)\ell i}\varphi_f
\varphi_\ell (N^2 - N_g N^g) \right],
\nonumber\\
&&\label{first}
\end{eqnarray}
where this last expression should be understood to be symmetrized
over both $pq$ {\it and} $ij$. Finally
\begin{eqnarray}
\{\Phi^{(\lambda)}_{0i} , \kappa^{(\lambda)}_{0j} \} &=&
\frac{N}{\sqrt{\gamma}} \gamma_{ij} \left[
\left(\frac{\pi_N}{N \lambda^{00}} \right)N_f + \varphi_f \right]^2,
\nonumber
\\
\{\Phi^{(\lambda)}_{0i} , \kappa^{(\lambda)}_{mn} \} &=&
\frac{\pi_N}{2 \sqrt{\gamma} \lambda^{00}} \left[
\left(\frac{\pi_N}{N \lambda^{00}} \right)N^f
+ \varphi^f \right]\left[ \gamma_{mi} \gamma_{nf}
+ \gamma_{ni}\gamma_{fm} - \gamma_{if}\gamma_{mn} \right],
\nonumber
\\
\{\Phi^{(\lambda)}_{ij} , \kappa^{(\lambda)}_{mn} \} &=&
\frac{\pi_N^2}{4 N \sqrt{\gamma} (\lambda^{00})^2}
\left[ \gamma_{mi} \gamma_{nj} + \gamma_{mj}\gamma_{ni}
- \gamma_{ij}\gamma_{mn} \right].
\label{last}
\end{eqnarray}

{}From the $19\times 19$ matrix discussed in Sec.~\ref{Sec3C},
the following linear combination of primary constraints (denoted
by $\tilde{\Phi}^{00}_{(s)}$) commutes with all the $\{
\Phi_{\tilde{P}}, \kappa_{\tilde{P}} \}$:
\begin{eqnarray}
\tilde{\Phi}^{00}_{(s)} &=& \left[N^4 - 2N^2 \varphi_j^2
\pi^\lambda(\pi^\lambda - N^k \varphi_k) + \pi^\lambda
(\pi^\lambda - N^k \varphi_k)(\pi^\lambda
-2 N^\ell \varphi_\ell)\right] \Phi^{00}_{(s)}
\nonumber
\\
&-& \left[ N^2 \varphi_j^2 \pi^\lambda(\pi^\lambda - N^k \varphi_k)
+ N^j \varphi_j(N^2 \varphi_j^2 \pi^\lambda(\pi^\lambda
- N^k \varphi_k)^2\right] \varphi_i \Phi^{0i}_{(s)}
\nonumber
\\
&-& (\pi^\lambda - N^k \varphi_k)^2 \varphi_i \Phi^{ij}_{(s)}
\varphi_j - 2N^2(\varphi^i \Phi_{ij}^{(\lambda)} \varphi^j
- \varphi_i^2 \Phi^{(\lambda) k}{}_k ).
\label{D:tildePhi}
\end{eqnarray}
We have also proved that it commutes with $\Phi_N$ as well as
${\cal H}_0$. The secondary constraint
$\tilde{\kappa}_{00}^{(s)}$ which commutes with all the
$\{\Phi_{\tilde{P}} \}$ as well as $\Phi_N$ is given by the same
expression but where, on the right hand side, the primary
constraints $\Phi$ are replaced by secondary constraints $\kappa$
(with the same labels). One must also complement this
$\tilde{\kappa}_{00}^{(s)}$ with a linear combination of primary
constraints so that it commutes with all secondary constraints
[see our discussion below Eq.~(\ref{M18})].


\end{document}